\documentclass[pre,floatfix, superscriptaddress,twocolumn]{revtex4}
\usepackage{amsmath,bm,graphicx}
\usepackage{amssymb}
\usepackage{amsfonts}
\usepackage{graphicx}
\usepackage{epstopdf}
\usepackage{amssymb}
\usepackage{mathrsfs}
\usepackage{amsmath}
\usepackage{color}
\usepackage{latexsym}
\usepackage{amsfonts}
\usepackage{hyperref}
\usepackage{subfigure}
\usepackage{wasysym}
\usepackage{dcolumn}
\usepackage{bm}
\usepackage{subfigure}
\usepackage{natbib}
\usepackage{ulem}

\begin{document}
\title{Non-monotonous polymer translocation time across corrugated channels: comparison between Fick-Jacobs approximation and numerical simulations }

\author{Valentino Bianco}
\thanks{Authors equally contributed to this work\\
        valentino.bianco@univie.ac.at\\ 
        malgaretti@is.mpg.de}
\affiliation{Faculty of Physics, Universit\"at Wien, Sensengasse 8, 1090, Vienna, Austria.}
\affiliation{Departament de Fisica Fonamental, Universitat de Barcelona, Barcelona, Spain}

\author{Paolo Malgaretti}
\thanks{Authors equally contributed to this work\\
        valentino.bianco@univie.ac.at\\ 
        malgaretti@is.mpg.de}
\affiliation{Max-Planck-Institut f\"{u}r Intelligente Systeme, Heisenbergstr. 3, D-70569
Stuttgart, Germany}
\affiliation{IV. Institut f\"ur Theoretische Physik, Universit\"{a}t Stuttgart,
Pfaffenwaldring 57, D-70569 Stuttgart, Germany}
\affiliation{Departament de Fisica Fonamental, Universitat de Barcelona, Barcelona, Spain}

\date{\today}

\begin{abstract}
We study the translocation of polymers across varying--section channels. Using systematic approximations, we derive a simplified model that reduces the problem of polymer translocation through varying--section channels to that of a point--like particle under the action of an effective potential. 
Such a model allows us to identify the relevant parameters controlling the polymers dynamics and, in  particular, their translocation time. 
By comparing our analytical results with numerical simulations we show that, under suitable conditions, our model provides reliable predictions of the dynamics of both Gaussian and self--avoiding polymers, in two-- and three--dimensional confinement. 
Moreover, both theoretical predictions, as well Brownian dynamic results, show a non--monotonous dependence of polymer translocation velocity as a function of polymer size, a feature that can be exploited for polymer separation. 
\end{abstract}

\maketitle

\section{Introduction}
Many biological processes such as viral injection of DNA into host cells~\cite{Cacciuto2006,Marenduzzo2013}, DNA transport through membrane or organelles~\cite{Albers} and gene transferring between bacteria~\cite{ChenScience2005, Guo2014853} involve the translocation of bio--polymer through nano--channels and nano--pores. Moreover, due to technological applications such as polymer separation, DNA sequencing and protein sensing, polymer translocation phenomena has been largely investigated in recent years~\cite{KasianowiczPNAS1996,hanPRL1999,liNATUREMAT2003, dekkarNATURENANO2007,Muthukumar2008,Binder2010, LiuSCIENCE2010,cohenPRL2011, luoJCP2011, guoJCP2011, cohenJCP2012,aldao2012, zhangJCP2013,Fazli2013,Fazli2015}.

Up to now, much attention has been payed to the case of polymers translocating through pores or channels whose half--section $h$, is typically comparable to the monomer size $r_0$, and therefore is much smaller that the polymer gyration radius  (see for example the recent reviews on the topic~\cite{Milchev2010,MuthukumarChapter,Jung2015}). 
In contrast to this, we are concerned with the case of a polymer translocating  
through channels, whose cross--section is larger than the size of a single monomer, yet comparable to the size of the overall gyration radius of the polymer. In such a regime,  the bottleneck can accommodate more than a single monomer at a time. This means, in principle, that the polymer can cross the  
bottleneck in a number of structural configurations where the first passing monomer is not necessary the ``head'' of the polymer.
Therefore, in the regime under study, the translocation dynamics of polymers reminds that of a deformable object through a constriction rather than threading through the eye of the needle.

\begin{figure}[t]
 \includegraphics[scale=0.37]{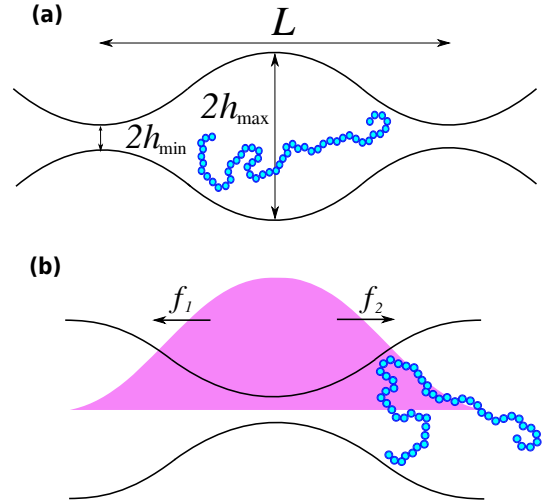}
 \caption{(a) schematic representation of a polymer moving inside a varying--section channel characterized by a maximum and minimum width, $h_{\rm max}$ and $h_{\rm min}$ respectively, and by a channel period $L$. (b) schematic representation of the free energy barrier (pink region) experienced by the polymer embedded in a varying--section channel. $f_1$ and $f_2$ are the effective forces acting on the polymer on both halves of the channel.}
\label{channel}
\end{figure}

In the following, we show that the regime $h\gg r_0$ leads to novel scenarios absent for the case $h\simeq r_0$. In particular, we focus our attention on two main features. 
Firstly, we show that when $h\gg r_0$ and for channels whose half--section $h(x)$, is varying smoothly, $\partial_{x}h(x)\ll1$, it is possible to reduce the dynamics of confined polymers to that of their center of mass regarded as a  point--like particle moving in an effective potential. Moreover, we show that for $L/R_G\gg 1$, namely when the channel longitudinal size $L$, is much larger than the polymer gyration radius $R_G$, it is possible to derive the effective potential from the equilibrium free energy of a polymer confined between parallel plates. In this view, our model extends the well--known Fick--Jacobs approximation~\cite{Diffusion_Processes_Jacobs1935,zwanzig,RegueraPRE2001} to the case of polymer dynamics across corrugated channels. 
Secondly, we show the reliability and the limit of validity of our model by comparison with numerical simulations. Such a comparison shows that our model properly captures the translocation dynamics of polymers, both at equilibrium as well as under the action of mild forces. 
Finally, our model predicts a non--monotonic dependence of the translocation velocity on polymer size and it provides an insight into the physical origin of such a behavior. These predictions are confirmed by numerical simulations. 

The structure of the text is the following. In section II we derive our model, i.e. we extend the Fick--Jacobs equation to the case of polymers embedded in varying--section channel. In section III we present our numerical scheme. In section IV we compare our predictions to the numerical solution of the Langevin equation. In section V we summarize our results. 

\section{Theoretical framework}

\subsection{Fick-Jacobs approximation}
The staring point of our model is the Fick--Jacobs approximation~\cite{Diffusion_Processes_Jacobs1935,zwanzig,RegueraPRE2001} that has already been characterized~\cite{,Kalinay2006,Kalinay2008,hanggi_2011,dagdug2013,dagdug2013-2,dagdug2015,malgarettiFrontiersPhysics2013} and exploited for diverse systems ranging from particle splitters~\cite{Reguera2012,Motz2014}, cooperative rectification~\cite{Malgaretti2012,paolo_jcp_2013,paolo_working_conf} diffusion through porous media~\cite{Dagdug2012,Umberto2015}, electro--osmotic systems~\cite{PaoloElecotrokinetics,paolo_macromolecules,paolo_jcp_2015} and entropic stochastic resonance~\cite{Burada2008,Huai2015} just to mention a few cases among others. The Fick--Jacobs approximation allows us to project the convection--diffusion equation of a non--interacting particle, confined in a two--dimensional ($2\mathscr{D}$) or three--dimensional ($3\mathscr{D}$) corrugated channel, onto a one--dimensional ($1\mathscr{D}$) equation in which the particle dynamics is controlled by an effective potential. 
Such an approximation is accurate for smoothly--varying channels, 
i.e. when $\partial_x h(x)\ll 1$, and for mild values of the external longitudinal force $f_0$, $\beta f_0 L\lesssim1$, being $L$ the length of the channel, $1/\beta=k_B T$ the inverse thermal energy, $k_B$ the Boltzmann constant and $T$ the absolute temperature. In fact, in such a regime, the motion along the longitudinal direction is slow enough so that the transverse probability 
retains its equilibrium shape (see Ref.\cite{Diffusion_Processes_Jacobs1935,zwanzig,RegueraPRE2001} for a derivation of the Fick--Jacobs approximation, Ref.~\cite{Kalinay2006,Kalinay2008,hanggi_2011,dagdug2013,dagdug2013-2,dagdug2015} for a discussion of its limits and Ref.~\cite{malgarettiFrontiersPhysics2013} for a review on recent applications). Under such assumptions the motion of a point--like particle is described by the time--dependent probability distribution $P(x,t)$ given by
\begin{equation}
\partial_{t}P(x,t)=\partial_{x}\left[\beta DP(x,t)\partial_{x}\mathcal{F}(x)+D\partial_{x}P(x,t)\right]
\label{eq:Fick-Jacobs}
\end{equation}
where $D$ is the diffusion coefficient of the particle and 
\begin{equation}
 \mathcal{F}(x)\equiv -k_B T\ln\left[\int_{-\infty}^{\infty}dz \int_{-\infty}^{\infty} e^{-\beta W(x,y,z)}dy\right].
 \label{eq:FJ1}
\end{equation}
The potential
\begin{eqnarray}
 W(x,y,z)\equiv \begin{cases}
 \psi(x,y,z)-f_0 x, &\sqrt{y^2+z^2}\le h(x)\\
 \infty, &\sqrt{y^2+z^2}> h(x) 
 \end{cases}
\label{potential}
\end{eqnarray}
accounts for both the geometrical confinement and for other possible conservative potentials $\psi$, and external forces $f_0$ acting along the longitudinal axis of the channel. We stress that, for vanishing external force $f_0=0$, $\mathcal{F}(x)$ reduces to the local equilibrium free energy, since it is the logarithm of the local partition function $\mathcal{Z}(x)\equiv \int_{-\infty}^{\infty}dz \int_{-\infty}^{\infty} e^{-\beta W(x,y,z)}dy$. At steady state we can solve for $P(x,t)\equiv P(x)$ in Eq.~(\ref{eq:Fick-Jacobs}), getting
\begin{equation}
P(x)=e^{-\beta\mathcal{F}(x)}\left[-\dfrac{J}{D}\int_{-\frac{L}{2}}^xe^{\beta\mathcal{F}(z)}dz+\Pi\right]
\label{distr-prob-gen}
\end{equation}
where the flux $J$ and $\Pi$ are determined by the boundary conditions and the normalization of $P(x)$. In particular, at equilibrium $J=0$ and $P(x)$ reads
\begin{equation}
P(x)=\dfrac{1}{\mathcal{Z}}e^{-\beta\mathcal{F}(x)}
\label{prob_distr}
\end{equation}
with $\mathcal{Z}\equiv\int_{-\frac{L}{2}}^{\frac{L}{2}} e^{-\beta\mathcal{F}(x)}dx$. For $f_0\neq 0$ and periodic boundary conditions we have
\begin{equation}
 \Pi = -\dfrac{J}{D}\dfrac{e^{-\beta\mathcal{F}(\frac{L}{2})}\int_{-\frac{L}{2}}^{\frac{L}{2}}e^{\beta\mathcal{F}(x)}dx}{e^{-\beta\mathcal{F}(-\frac{L}{2})}-e^{-\beta\mathcal{F}(\frac{L}{2})}}=-\dfrac{J}{D}\Pi_0
 \label{eq:polymer-flux-Pi},
\end{equation}
and the flux reads
\begin{equation}
 J = -D\left[\int_{-\frac{L}{2}}^{\frac{L}{2}}dx e^{-\beta\mathcal{F}(x)}\left(\int_{-\frac{L}{2}}^{x}e^{\beta\mathcal{F}(z)}dz+\Pi_0\right)\right]^{-1},
 \label{eq:polymer-flux-a}
\end{equation}
from which we can define the translocation velocity
\begin{equation}
v=J L.
\label{eq:def-mu}
\end{equation}
According to Eqs.~(\ref{eq:FJ1}) and (\ref{eq:def-mu}), the translocation velocity $v$ is determined once a local expression for $ \mathcal{F}(x)$ is known. 

\subsection{Gaussian polymer}

The partition function of a Gaussian polymer that undergoes solely steric interaction with the walls of a box of sizes $L_{x}$, $L_{y}$ and $L_{z}$ factorizes~\cite{doi_edwards}

\begin{equation}
 Z=Z_{x}Z_{y}Z_{z},
\end{equation}
being each term of the form
\begin{equation}
Z_{i}=\frac{8}{\pi^{2}}L_{i}\sum_{p=1,3,..}^{\infty}\frac{1}{p^{2}}\exp\left(-\frac{\pi^{2}p^{2}}{L_{i}^{2}}R_G^2\right),
\label{eq:doi-edw-part-funct}
\end{equation}
where $R_G=(Nb^2/6)^{1/2}$ is the radius of gyration of a Gaussian polymer in an unbound medium~\cite{doi_edwards}, $b$ is the equilibrium distance between contiguous monomers and $N$ is the number of monomers. 
In principle, Eq.~(\ref{eq:doi-edw-part-funct}) is valid for polymer embedded between planar walls. However, for $L\gg R_G$, namely when the channel sections varies on length scales much larger than $R_G$, we can 
approximate the local partition function of a polymer, whose center of mass is located at position $x$, by Eq.~(\ref{eq:doi-edw-part-funct}) evaluated for $L_{i}=2h(x)$. 
In the following, we  consider a polymer embedded in a $2\mathscr{D}$ or $3\mathscr{D}$ channel. In both cases, we assume that the channel is symmetric with respect to its longitudinal $x$--axis. 
Accordingly, the local equilibrium free energy $\mathcal{F}_{eq}$ for a confined Gaussian polymer 
reads
\begin{multline}
\mathcal{F}_{eq}(x)=-\frac{1}{\beta}(d-1)\ln\Biggl[\frac{16h(x)}{h_0\pi^{2}}\times\\
\times\sum_{p=1,3,..}^{\infty}\frac{1}{p^{2}}\exp\left(-\frac{\pi^{2}p^{2}}{4h^{2}(x)}R_G^2\right)\Biggl],
\label{eq:free-energy1}
\end{multline}
where $h_0$ is the average channel section and $d=2$, $d=3$ refer to $2\mathscr{D}$ and $3\mathscr{D}$ channel respectively. 
We can separate $\mathcal{F}_{eq}$ into two contributions by rewriting Eq.~(\ref{eq:free-energy1}) as
\begin{multline}
\mathcal{F}_{eq}(x)=-\frac{1}{\beta}(d-1)\Biggl\{\ln\left[\frac{16h(x)}{h_0\pi^{2}}\right]+\\
+\ln\left[\sum_{p=1,3,..}^{\infty}\frac{1}{p^{2}}\exp\left(-\frac{\pi^{2}p^{2}}{4h^{2}(x)}R_G^2\right)\right]\Biggl\}.
\label{eq:free-energy1b}
\end{multline}
The first, $\ln\left[16h(x)/h_0\pi^{2}\right]$, can be identified  as the entropy of the center of mass of the polymer, regarded as a point--like particle embedded in a varying--section channel with half--section $h(x)$~\cite{zwanzig,RegueraPRE2001}. The second term accounts for the correction due to the presence of the other monomers. The two terms in Eq.~(\ref{eq:free-energy1b}) identify two regimes. For polymers whose size is smaller compared to the channel width, namely $R_G/h(x)\ll 1$, the second term in Eq.~(\ref{eq:free-energy1b}) can be considered as constant and $\mathcal{F}_{eq}(x)$ becomes independent on the polymer properties, being equivalent to the a point--like particle~\cite{RegueraPRE2001,zwanzig}. On the contrary, for $R_G/h(x)\gg 1$ the second term in Eq.~(\ref{eq:free-energy1b}) is dominating and the overall dynamics depends on the polymer properties encoded in $R_G$.

Eq.~(\ref{eq:free-energy1b}) is modified by an external field $f_0$ acting on all $N$ monomers along the longitudinal axis of the channel. 
Following the Fick--Jacobs approximation we assume that, in the regime $\partial_x h(x)\ll 1$ and for mild external forces, $\beta f_0 L\lesssim 1$, the local free energy can be written as 
\begin{multline}
\mathcal{F}(x)=-Nf_{0}x-\frac{1}{\beta}(d-1)\Biggl\{\ln\left[\frac{16h(x)}{h_0\pi^{2}}\right]+\\
\ln\left[\sum_{p=1,3,..}^{\infty}\frac{1}{p^{2}}\exp\left(-\pi^{2}p^{2}\left(\frac{R_G}{2h(x)}\right)^2\right)\right]\Biggl\}.
\label{eq:free-energy1c}
\end{multline}
Eq.~(\ref{eq:free-energy1c}) accounts for the contributions to the local free energy due to all the accessible polymer configurations and represents the effective potential experienced by the center of mass of the polymer. 

\begin{figure*}[t]
\centering
\includegraphics[scale=0.9]{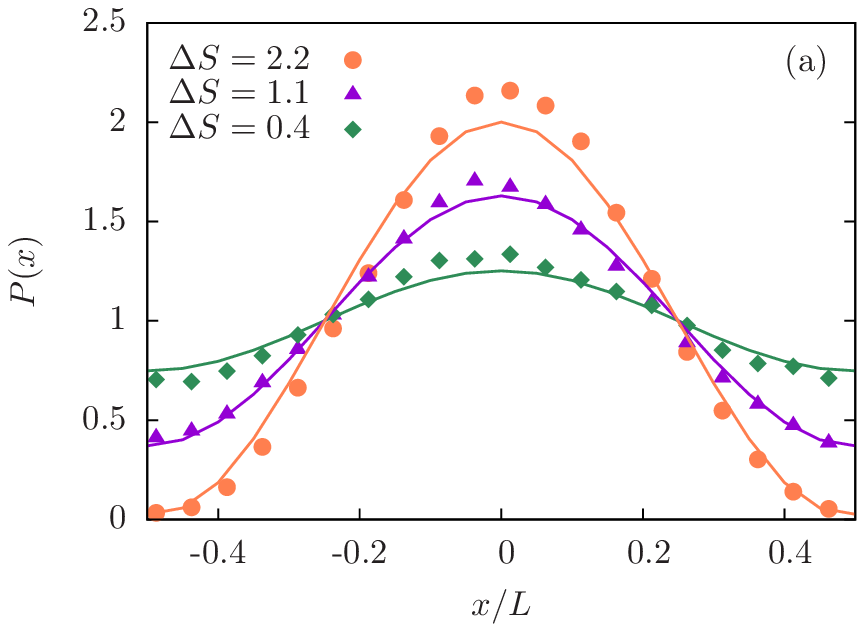}
\includegraphics[scale=0.9]{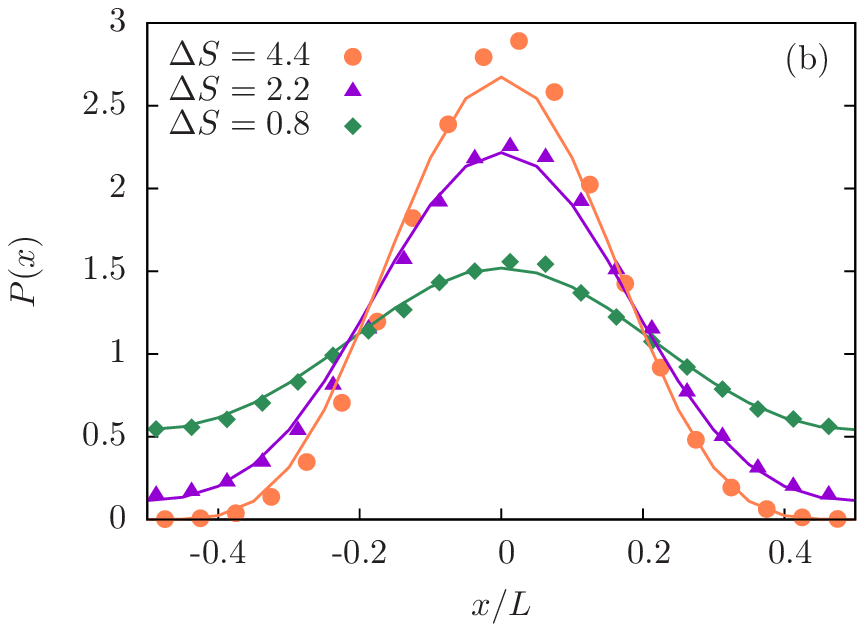}\\
\vspace{0.4 cm}
\caption{Theoretical (lines) and numerical (dots) values  of the equilibrium probability distribution $P(x)$, as defined in Eq.~(\ref{prob_distr}). Panels (a) and (b) show $P(x)$ for $2\mathscr{D}$ and $3\mathscr{D}$ confinement, respectively. Data refer to a $N=20$ Gaussian polymer and different entropic barriers $\Delta S$.}
\label{fig:prob_distr-eq}
\end{figure*}

\subsection{Self--avoiding polymers}

For the case of a confined self--avoiding polymer there is no (to the best of our knowledge) explicit analytical expression for the free energy. However, we can exploit a scaling argument of De Gennes~\cite{DeGennes_book} according to which the free energy difference between two points of the channel with different width $h(x)$ should scale as 
\begin{equation}
 \beta\Delta\mathcal{F}\simeq \left[\frac{R_G}{h(x)}\right]^{\frac{1}{\nu}},
\end{equation}
where $\nu=3/(d+2)$ is the Flory exponent for self--avoiding polymers~\cite{doi_edwards}.
Even though such a scaling is valid only for the strong confinement regime $h_0\ll R_G$~\cite{DeGennes_book,Vliet1992}, we assume it to hold also in the weak confinement regime $h_0\sim R_G$, checking the validity of such an assumption by comparing  the theoretical predictions with numerical simulations. Accordingly, we assume that the local free energy of a self--avoiding polymer confined in section--varying channel, having its center of mass at position $x$, reads
\begin{multline}
\mathcal{F}(x)=-Nf_{0}x-\frac{1}{\beta}(d-1)\Biggl\{\ln\left[\frac{16h(x)}{h_0\pi^{2}}\right]+\\
+\ln\left[\sum_{p=1,3,..}^{\infty}\frac{1}{p^{2}}\exp\left(-\pi^{2}p^{2}\left(\frac{R_G}{2h(x)}\right)^{\frac{1}{\nu}}\right)\right]\Biggl\},
\label{eq:free-energy2}
\end{multline}
that, at leading order in $R_G$, leads to $\beta\mathcal{F}\simeq [R_G/2h(x)]^{1/\nu}$.

\begin{figure*}[t]
\includegraphics[scale=0.9]{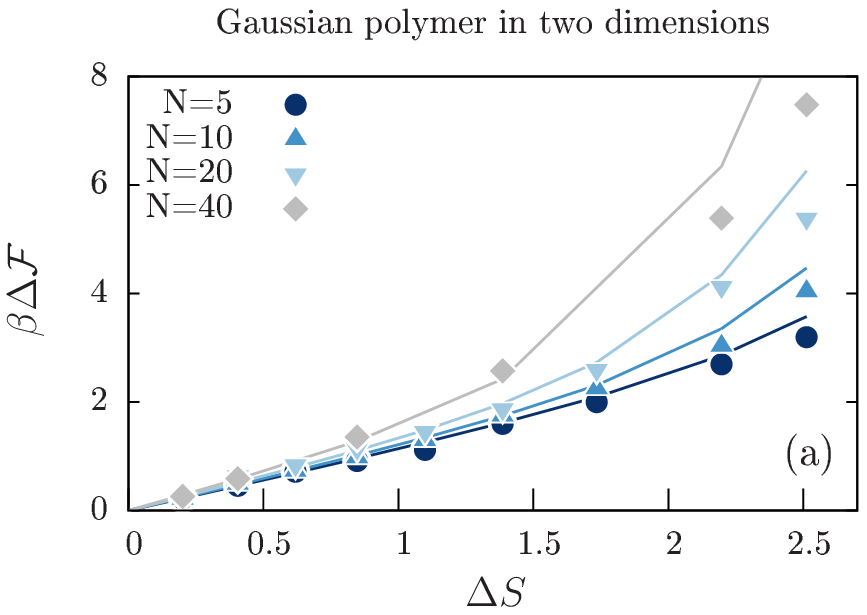}
\includegraphics[scale=0.9]{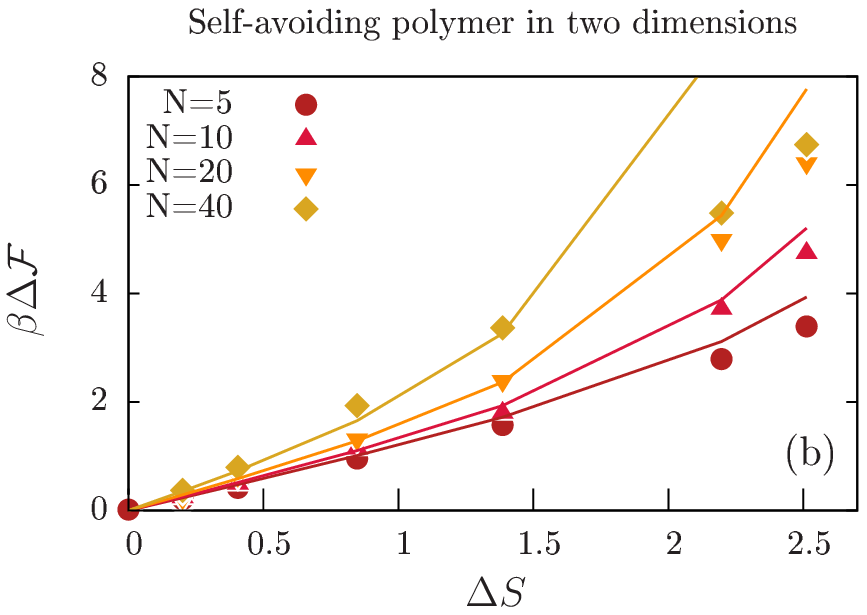}\\
\includegraphics[scale=0.9]{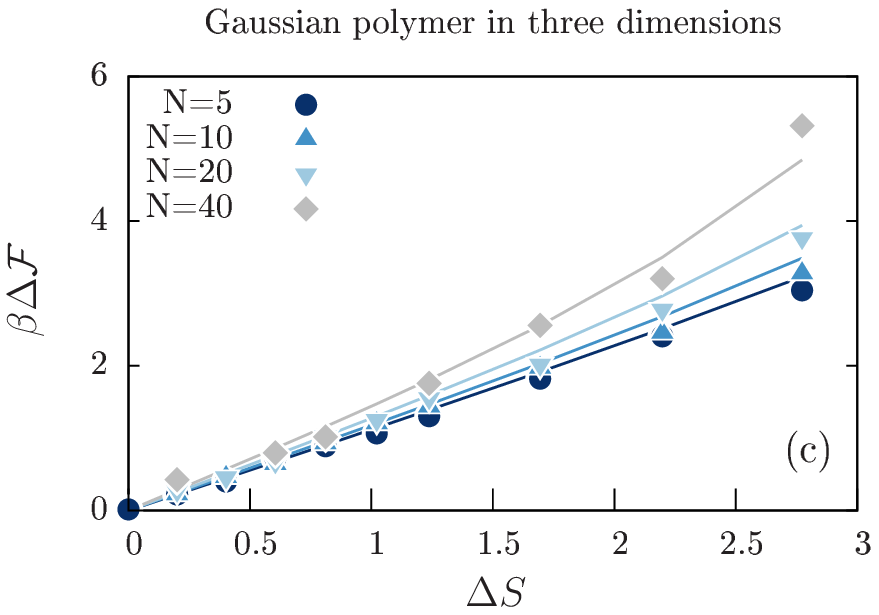}
\includegraphics[scale=0.9]{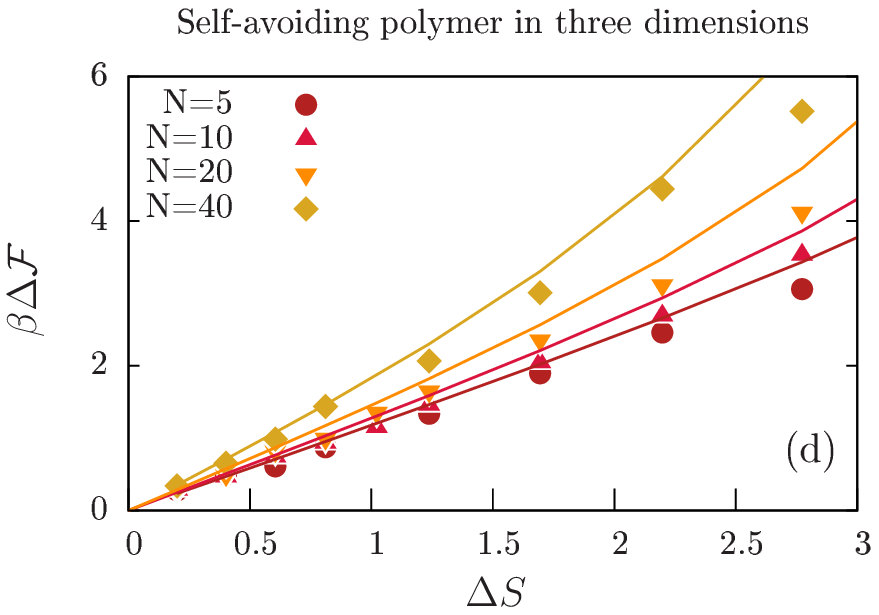}
\caption{Equilibrium ($f_0=0$) free energy difference $\Delta\mathcal{F}_{eq}$ for Gaussian (left panels) and self--avoiding (right panels) polymers of different sizes as a function of $\Delta S$. Data from simulations, obtained by averaging over hundreds of translocation events, are plotted with points while lines are the theoretical predictions: Eq.~(\ref{eq:free-energy1c}) for Gaussian polymers shown in panel (a) for $2\mathscr{D}$ channel and panel (c) for $3\mathscr{D}$ channel;  Eq.~(\ref{eq:free-energy2}) for self--avoiding polymers shown in panel (b) for $2\mathscr{D}$ channel (for which $\nu = 3/4$ in Eq.~(\ref{eq:free-energy2})) and panel (d) for $3\mathscr{D}$ channel (for which $\nu = 3/5$ in Eq.~(\ref{eq:free-energy2})). }
\label{fig:eq-fig}
\end{figure*}

\subsection{Regimes of validity}
We expect Eq.~(\ref{eq:free-energy1c}), or equivalently Eq.~(\ref{eq:free-energy2}), to properly capture the dynamics of confined polymers when polymers translocation is slow enough so that polymers can explore the transverse directions according to the equilibrium Boltzmann weight.

At equilibrium, the system is characterized by two time scales: i) the diffusion time $\tau_{D}=\frac{L^{2}}{D_N}$ that the  center of mass of the polymer takes to diffuse across the channel; ii) the relaxation time $\tau_{R}=\frac{2R_G^{2}}{\pi^{2}D_N}$ of the slowest Rouse mode (the end--to--end mode). According to the Rouse model~\cite{doi_edwards}, the diffusion coefficient is $D_N=\frac{D_0}{N}$, being $D_0$ the diffusion coefficient of a single monomer. 
Hence, at equilibrium, our approximation holds for $\dfrac{\tau_{D}}{\tau_{R}}\gg 1$ that reads
\begin{equation}
R_G\ll \frac{\pi}{\sqrt{2}}L,
\label{eq:check1}
\end{equation}
implying that the channel length $L$ must be bigger than the gyration radius of the polymer $R_G$.

When the system is under the action of an homogeneous external force $f_0$, the translocation time $\tau_{F}=\frac{L}{\beta D_N N f_0}$ due to the advection by the external force appears as an additional time scale. For weak forces, $\beta D_N N f_0 \ll D_N/L$, we have $\tau_D<\tau_F$ and therefore the validity of our approximation is still controlled by Eq.~(\ref{eq:check1}). On the contrary, for $\beta D_N N f_0 \gg D_N/L$ we have $\tau_D>\tau_F$ and imposing $\frac{\tau_F}{\tau_R}\gg 1$ leads to
\begin{equation}
 \frac{\beta D_N N f_0}{L} \ll \frac{\pi^{2}D_N}{2R_G^{2}}.
 \label{eq:def-reg-val}
\end{equation}

\section{Simulation details}

In order to check if Eq.~(\ref{eq:free-energy1c}) and Eq.~(\ref{eq:free-energy2}), via Eq.~(\ref{eq:polymer-flux-a}), provide a reliable 
description of polymer dynamics, 
we perform Brownian dynamics simulations (over--damped regime) of a $N$ monomers bead--spring polymer confined in a channel, for $d=2,3$, for both cases of a Gaussian and self--avoiding polymer. Such a polymer model is widely used (see for example Ref.~\cite{zhangJCP2012, AguilarSoftMatter2012, XingEPL2014, YamazakiJCPB2014, DebashishNatureComm2014} just to mention some recent works).
The interaction between neighbor monomers is given by an harmonic potential
\begin{equation}\label{armonic}
 U^{\rm harm}_{\rm nn}=-\frac{K}{2} \left(|{\bf r}_i - {\bf r}_{i+1}| - b \right)^2 ,
\end{equation} 
where $b$ is the radius of a monomer and $i=1\ldots N-1$ \footnote{A Gaussian chain can be represented as a mechanical model on $N$ beads connected by harmonic springs according to Ref.~\cite{doi_edwards}.}. 
In the case of the
self--avoiding polymer, the repulsive interaction between non--contiguous monomers is accounted by a repulsive harmonic potential acting among all beads 
\begin{equation}\label{armonic1}
\begin{cases}
 U^{\rm harm}_{\rm rep}=-\frac{K}{2} \left(|{\bf r}_i - {\bf r}_j| - b  \right)^2 & \text{for  }
|{\bf r}_i - {\bf r}_j| < b\\\\
U^{\rm harm}_{\rm rep}=0 & \text{for  } |{\bf r}_i - {\bf r}_j| \geq b\
\end{cases},
\end{equation}
with $|i-j|\geq 2$. We fix $K=100$ $k_BT/r_0$ to prevent the crossing events between polymer bonds.
The interaction with the channel surface is modeled with an infinite wall potential $U^{\rm surf}$
\begin{equation}\label{wall}
\begin{cases}
 U^{\rm surf}=\infty \text{ }\text{ for }\text{ } h(x)>h_0+h_1\cos(2\pi x/L)
 \\\\
U^{\rm surf}=0\qquad \text{ otherwise}
\end{cases},
\end{equation}
and reflecting condition are applied when a monomer attempt to cross the wall~\footnote{Following Ref.~\cite{Motz2014}, if during the integration step a monomer crosses the channel surface the move is rejected.}. Finally, a net constant force ${\bf F}$ along the channel axis is applied to each monomer.
The polymer is embedded in periodic channel, with half width described by
\begin{equation}
h(x)=h_{0}+h_{1}\cos(2\pi x/L).
\end{equation}
The Euler algorithm is used to integrate the equation of motion 
\begin{equation}\label{eq_brownian}
{\bf\dot{r}}=\beta D_0\left ( -\boldsymbol{\nabla} U + {\bf F} + \boldsymbol{\eta} \right ),
 \end{equation} 
where $U=U^{\rm arm}(r) + U^{\rm surf}(r)$ is the total potential, ${\bf \eta}$ is the random force accounting for the monomer--solvent interaction and satisfying the fluctuation--dissipation relation $\langle {\bf\eta}_i(t){\bf \eta}_j(t')\rangle = 2dk_BT\zeta\delta_{ij}\delta(t-t')$ and $\zeta=k_BT/D_0$ is the friction coefficient of a monomer. 
We adopt $k_BT$ and $r_0$ as unit of energy and distance and $r_0^2/D_0$ as unit of time. 
According to this units, we fix the integration time step of Eq.~(\ref{eq_brownian}) to $dt=10^{-3}$, $L=40$ and $h_0=10$.

\section{Results and discussions}

\subsection{Equilibrium}

As a first check of the reliability of our model we have controlled if it properly recovers equilibrium properties of the system such as the equilibrium probability distribution function $P(x)$ of the center of mass of the polymer as a function of its longitudinal position (see Eq.~(\ref{prob_distr})). 
Fig.~(\ref{fig:prob_distr-eq}) shows the good agreement between the analytical predictions of $P(x)$ (Eqs. (\ref{prob_distr}) and (\ref{eq:free-energy1c})) and the outcome of Brownian dynamics simulations.
\begin{figure*}[t]
\centering
\includegraphics[scale=0.9]{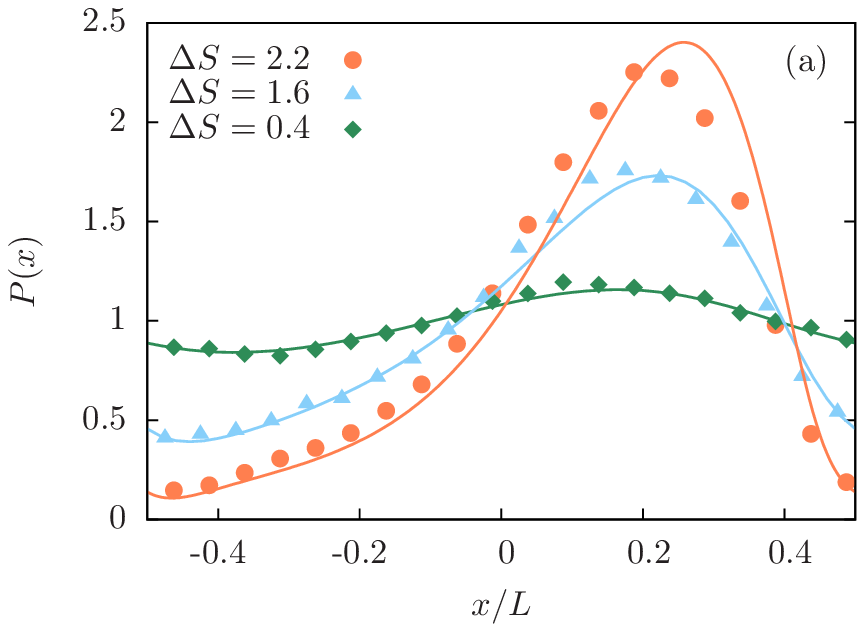}
\includegraphics[scale=0.9]{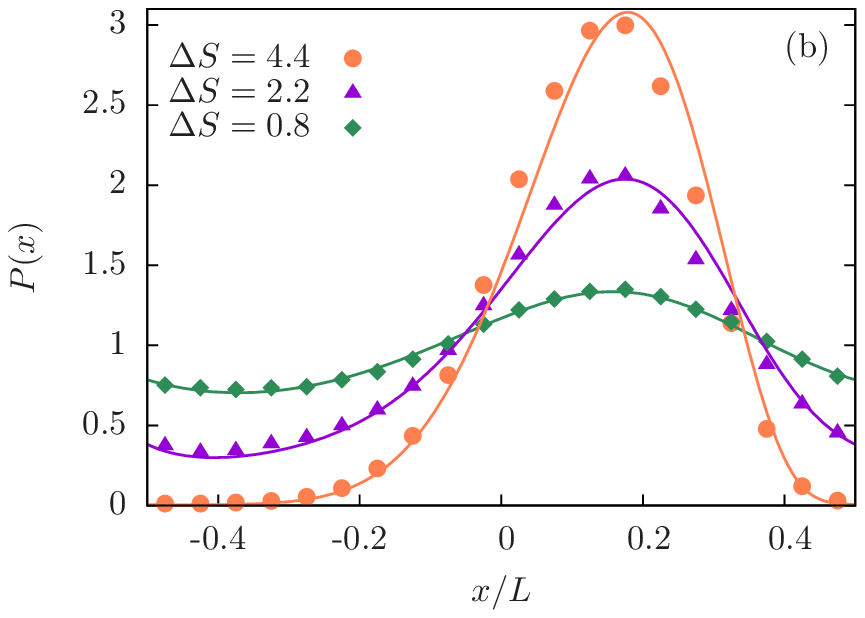}
\caption{Theoretical and numerical probability distribution $P(x)$ out of equilibrium, as defined in Eq.~(\ref{distr-prob-gen}), for a $N=20$ Gaussian polymer and different values of the entropic barrier $\Delta S$. Panels (a) and (b) show $P(x)$ for $2\mathscr{D}$ and $3\mathscr{D}$ confinement, respectively. The external force is $\beta f_0 L=0.4$.}
\label{fig:prob_distr-frc}
\end{figure*}
In order to quantify the mismatch between the theoretical prediction and the outcome of numerical simulations we  focus on the free energy difference 
\begin{equation}
\Delta\mathcal{F}_{eq}\equiv \ln P(0) - \ln P(L/2) 
\end{equation}
experienced by the Gaussian/self--avoiding polymers between the channel maximum amplitude (located at $x=0$) and the bottleneck (located at $x=\pm L/2$) for different values of $N$ and different channel geometries, encoded by the entropic barrier
\begin{equation}
 \Delta S\equiv (d-1)\ln\left(\frac{h_{\rm max}}{h_{\rm min}}\right),
\end{equation}
where $d=2,3$ stands for the $2\mathscr{D}$ and $3\mathscr{D}$ case respectively.

Fig.~\ref{fig:eq-fig} shows the numerical values of $\Delta\mathcal{F}_{eq}$, calculated by Brownian dynamics simulations (dots), along with the analytical prediction (lines) given by Eq.~(\ref{eq:free-energy1c}),(\ref{eq:free-energy2}) as a function of $\Delta S$. 
As shown in Fig.~\ref{fig:eq-fig}(a),(c), for Gaussian polymers our model fits remarkably well with the data obtained for small entropic barriers, up to $\Delta S \lesssim 2$, whereas the quantitative agreement weakens for larger values of $\Delta S$. 
The good agreement between theoretical and numerical predictions can be improved considering that, due to the confinement, the radius of gyration $R_G$ is not constant, rather it depends on the longitudinal position $x$. Accordingly, by substituting  the expression for $R_G(x)$ provided by Ref.~\cite{DebasishPRE} into Eq.~(\ref{eq:free-energy1c}) the agreement between the analytical predictions and the outcome of Brownian dynamics simulations improves (see Appendix A).
Surprisingly, Eq.~(\ref{eq:free-energy2}) works quite well for the self--avoiding polymer, as shown in Fig.~\ref{fig:eq-fig}(b),(d) and the range of values for which Eq.~(\ref{eq:free-energy2}) properly captures the free energy difference is similar to that observed for Gaussian polymers.

\subsection{External force}

\begin{figure*}[t]
\includegraphics[scale=0.9]{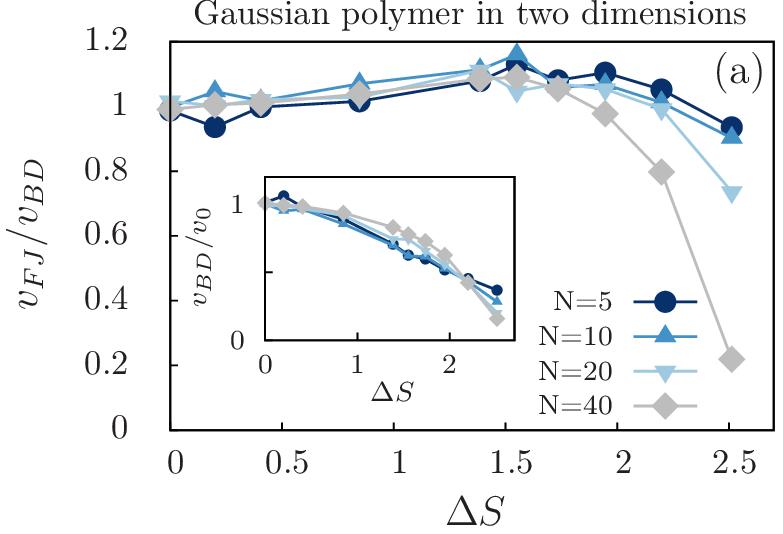}
\includegraphics[scale=0.9]{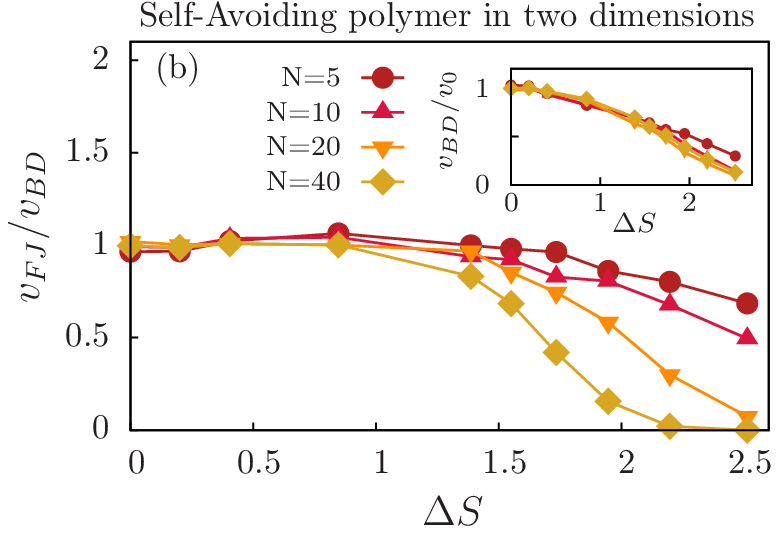}\\
\vspace{0.3 cm}
\includegraphics[scale=0.9]{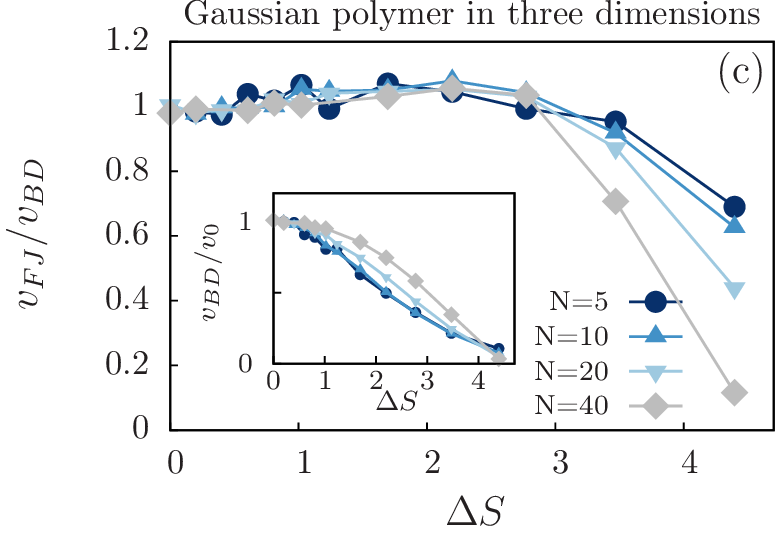}
\includegraphics[scale=0.9]{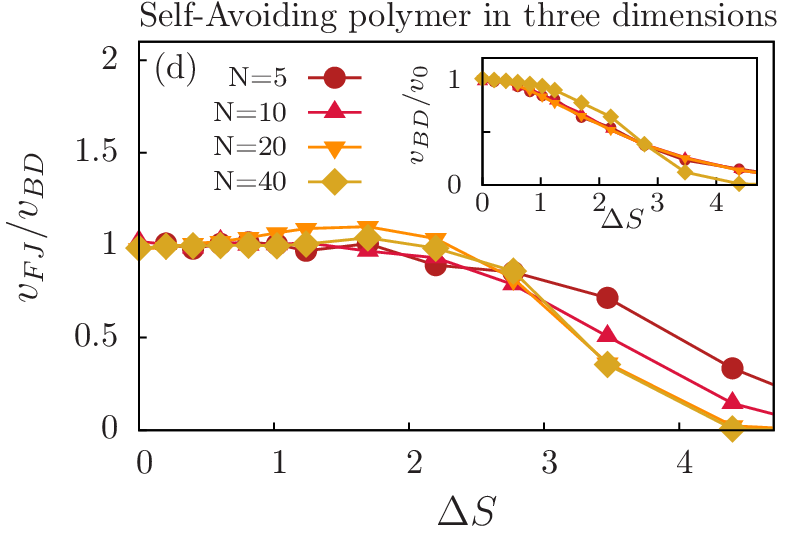}
\caption{Ratio between the theoretical prediction of the translocation velocity velocity $v_{FJ}$ (as defined in Eq.~(\ref{eq:def-mu})) and the velocity $v_{BD}$ calculated from Brownian dynamics simulations by averaging over hundreds of translocation events as a function of the entropic barrier $\Delta S$ for external force $\beta f_0 L=0.4$.
Panel (a) and (c) show the translocation velocity of a Gaussian polymer in $2\mathscr{D}$ and $3\mathscr{D}$ confinement respectively, whereas panel (b) and (d) show the translocation velocity of a self--avoiding polymer in $2\mathscr{D}$ (with $\nu=3/4$ in Eq.~(\ref{eq:free-energy2})) and $3\mathscr{D}$ (with $\nu=3/5$ in Eq.~(\ref{eq:free-energy2})) confinement.
Insets: $v_{BD}$ normalized by the velocity in a flat channel $v_0=\beta D_N N f_0$.  Lines are guides for eyes.}
\label{velocity_numerical_theoretical_f001}
\end{figure*}

The presence of a constant external force $f_0$ acting homogeneously on all monomers modifies the steady state probability distribution along the longitudinal axis of the channel. 
The panels in Fig.~\ref{fig:prob_distr-frc} show the good agreement between the probability distribution of the center of mass of a Gaussian polymer calculated from the numerical simulations (points) and the one predicted from Eq.~(\ref{distr-prob-gen}) with periodic boundary conditions (lines). 
The asymmetric probability distribution profile, shown in Fig.~\ref{fig:prob_distr-frc}, leads to the onset of a steady state non--vanishing translocation velocity $v$ as defined in Eq.~(\ref{eq:def-mu}).
Accordingly, we compare the theoretical prediction for the velocity of the center of mass, as given by Eq.~(\ref{eq:def-mu}), with that obtained by averaging the velocity of all the beads obtained by numerical simulations as a function of both polymer size $N$ and $\Delta S$ (Fig. \ref{velocity_numerical_theoretical_f001}). 
For small external forces $\beta f_0 L=0.4$, we observe that, for a Gaussian polymer, our model holds for values of $\Delta S \lesssim 1.5$ in two dimensions, and for $\Delta S \lesssim 2.5$ in three dimensions as shown in Fig.~\ref{velocity_numerical_theoretical_f001}.(a)(c). Our predictions become less reliable upon increasing $\Delta S$. 
For small entropic barriers, $\Delta S \leq 2$ for $d=2$ and $\Delta S \leq 3$ for $d=3$, 
the discrepancy between the numerical and theoretical value of the velocity remains in the range $3\% \div 10\%$ and does not show an increasing trend with the polymer size $N$. Such a discrepancy is more than acceptable considering that also the numerical estimations of the translocation velocity are affected by few percentage errors.
For increasing values of $\Delta S$ the mismatch between the model and the numerical simulation increases. Finally, comparing Fig.~\ref{velocity_numerical_theoretical_f001} with Fig.~\ref{fig:eq-fig} we notice that the range of validity of our model, for $\beta f_0 L=0.4$, is consistent with the one observed for the equilibrium case.  

Concerning self--avoiding polymers, the agreement between our model and numerical simulations is surprisingly good if we consider that Eq.~(\ref{eq:free-energy2}) represents a rough approximation of the local free energy. In particular, the better agreement between our model and the numerical simulation for the $3\mathscr{D}$ case as compared to the $2\mathscr{D}$ case may be due to the dimension mismatch between the $1\mathscr{D}$ polymer and the $2\mathscr{D}-3\mathscr{D}$ environment. 
In fact, in a $3\mathscr{D}$ environment the typical configurations of a self--avoiding polymers are qualitatively similar to those of Gaussian polymer. Accordingly, by accounting the diverse exponent characterizing the Gaussian and self--avoiding polymers, our ansatz, Eq.~(\ref{eq:free-energy3}),  provides a reliable approximation to the exact free energy. On the contrary, in a $2\mathscr{D}$ environment a self--avoiding polymer experiences a stronger constraint on the available configurations. Therefore our ansatz, that accounts for the self--avoiding nature of the polymer solely via the diverse scaling exponents, fails to capture the underlying dynamics and our predictions becomes less reliable.

For larger values of $f_0$ ($\beta f_0 L=4$) the external force outnumbers the entropic contribution in the free energy induced by the confinement. This enhances the agreement between theoretical and numerical polymer translocation velocity. Such an agreement holds also for even larger entropic barriers, for both Gaussian as well as self--avoiding polymers (see Fig.~\ref{velocity_numerical_theoretical_f01} in Appendix B). According to Ref.~\cite{BuradaPRE2007} (in particular Eq. 26 therein) we expect that our approximation holds also for larger forces, up to $\beta f_0 L \sim 300$.

\begin{figure*}[t]
\includegraphics[scale=0.99]{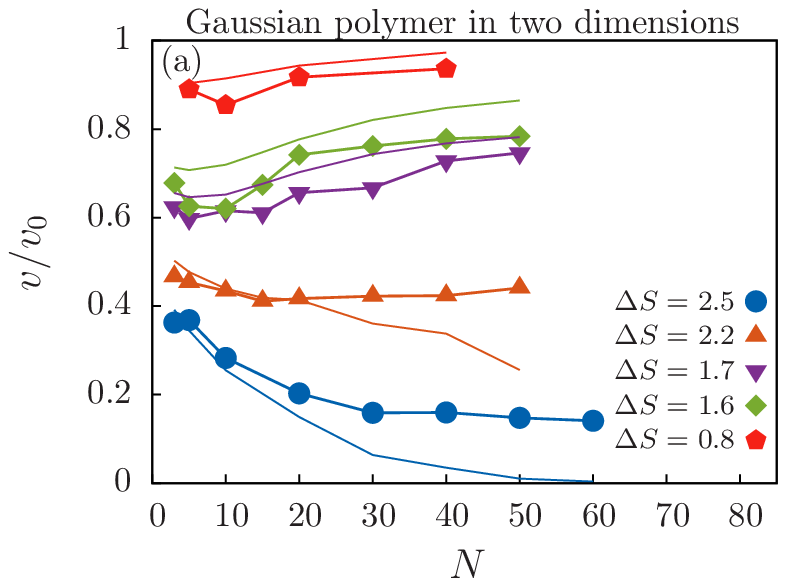}
\includegraphics[scale=0.99]{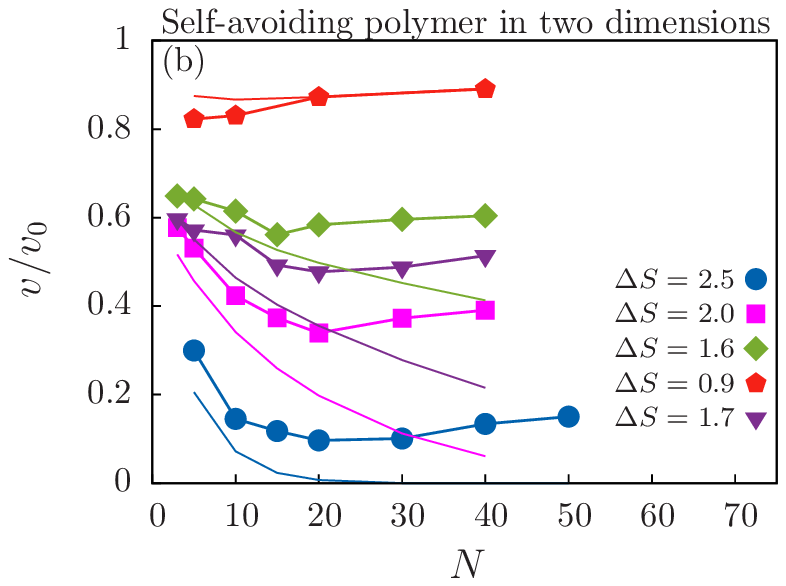}\\
\vspace{0.3 cm}
\includegraphics[scale=0.99]{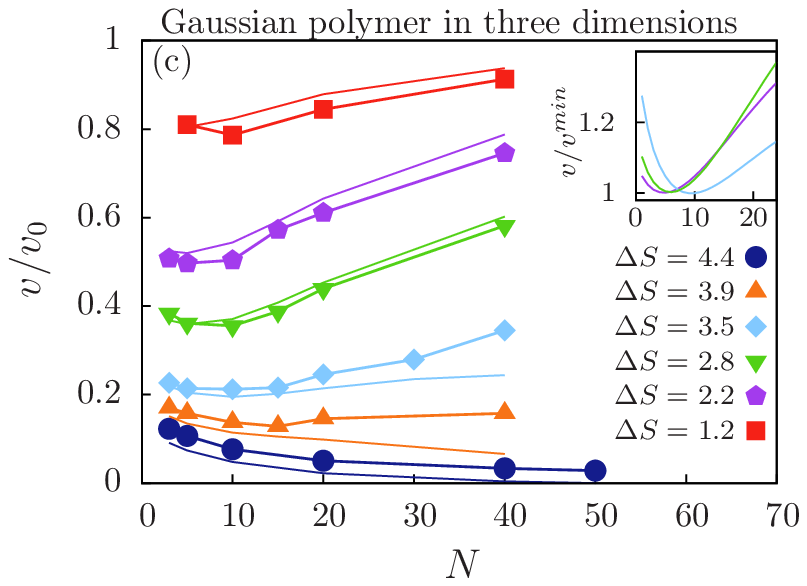}
\includegraphics[scale=0.99]{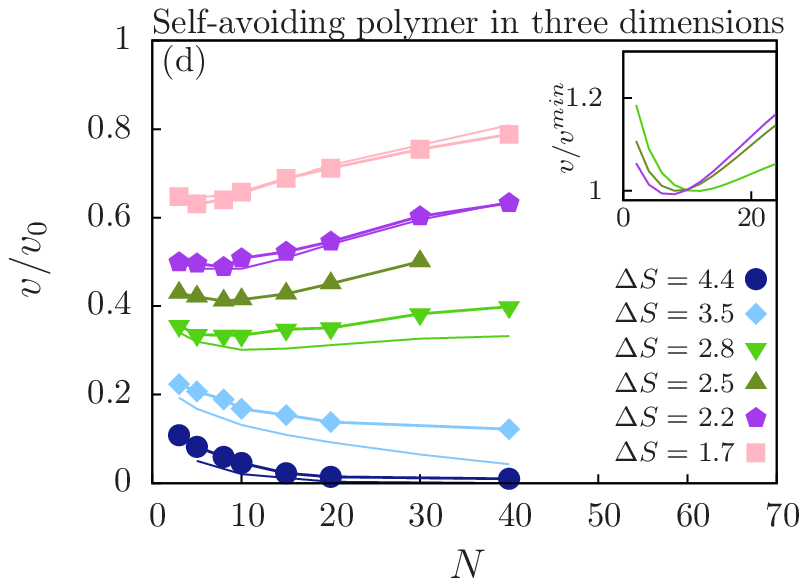}
\caption{Translocation velocity of Gaussian (left panels) and self--avoiding (right panels) polymers confined in two $2\mathscr{D}$ (upper panels) or $3\mathscr{D}$ (lower panels) varying--section channels under the action of a constant force $\beta f_0 L=0.4$ (acting on all monomers), as function of $N$. $v$ is normalized by the velocity $v_0=\beta D_0 f_0$ in a constant section channel.
Points represent the translocation velocities obtained from Brownian dynamics simulations by averaging over hundreds of translocation events, whereas lines represent the theoretical prediction calculated according to Eqs.~(\ref{eq:polymer-flux-a}),(\ref{eq:def-mu}), where the free energy is given by Eq.~(\ref{eq:free-energy1c}) and Eq.~(\ref{eq:free-energy2}) for Gaussian and self--avoiding polymers respectively. Numerical errors are of the order of symbol size.
The insets show the velocity from numerical predictions normalized by its minimum value. 
Lines connecting points are guides for eyes.}
\label{velocity_polymer_length_NO_SELF_f001_multipanel}
\end{figure*}

The dependence of polymer velocity $v$ on the polymer size $N$ shows an interesting behavior.
Firstly, Fig.~\ref{velocity_polymer_length_NO_SELF_f001_multipanel} shows that for $N\rightarrow \infty$ and a given values of $f_0$ the value of $v$ is determined by $\Delta S$. 
While larger values of $\Delta S$ lead to vanishing velocity of longer polymers, smaller values of $\Delta S$ lead to  an increase of the velocity that eventually saturates.
Secondly, for finite values of $N$ and for mild entropic barriers, namely $\Delta S\simeq 2$ for $d=2$ and $\Delta S \simeq 3$ for $d=3$, the velocity is not monotonous upon increasing $N$, rather it shows a minimum for both polymer models, and for $2\mathscr{D}$ and $3\mathscr{D}$ confinement (see Fig.~\ref{velocity_polymer_length_NO_SELF_f001_multipanel} and the corresponding insets).
Finally, Figs.~\ref{velocity_polymer_length_NO_SELF_f001_multipanel}(c),(d) show that, for intermediate values of $\Delta S$, Gaussian polymers are faster than equally long self--avoiding ones. For example, for $\Delta S=2.77$ and $N=40$ Gaussian polymers are almost twice as fast as self--avoiding polymers. Interestingly, the magnitude of the velocity difference between Gaussian and self--avoiding polymers can be tuned by varying $\Delta S$, opening the chance for a solvent--sensitive entropic separation mechanism for polymers. In fact, polymers for which the solvent is close to their $\Theta$ condition  behave more similar to Gaussian polymers, whereas polymers for which the solvent is a ``good'' solvent  behave as self--avoiding ones, hence leading to a solvent dependent velocity that can be exploited to separate polymers.

In order to understand the dependence of the asymptotic values of $v$ upon variation of $\Delta S$, as well as the physical origin of the non monotonous dependence of $v$ on $N$, we simplify Eq.~(\ref{eq:free-energy1c}) by accounting only for the first exponential term in the sum, i.e. we reduce the Eq.~(\ref{eq:free-energy1c}) to the \textit{ground state}~\cite{DeGennes_book}
\begin{equation}
 \mathcal{\tilde{F}}(x)=-Nf_{0}x+\frac{d-1}{\beta}\left[\ln\left(\frac{16h(x)}{h_0\pi^{2}}\right)-\pi^{2}\frac{R_G^2}{4h(x)^2}\right].
\label{eq:free-energy3}
\end{equation}
We further approximate $\mathcal{\tilde{F}}(x)$ by a piece-wise linear function
\begin{equation}
\mathcal{\tilde{F}}(x)\simeq\begin{cases}
-f_{1} x=-\left(Nf_0+\frac{2}{L}\Delta\mathcal{\tilde{F}}_{eq}\right)x, & -L/2<x\leq0\\\\
-f_{2} x=-\left(Nf_0-\frac{2}{L}\Delta\mathcal{\tilde{F}}_{eq}\right)x, & 0\leq x<L/2
\end{cases},
\label{eq:approx-free-en}
\end{equation}
with
\begin{equation}
  \Delta\mathcal{\tilde{F}}_{eq} \equiv
  -\frac{d-1}{\beta}\left\{\ln\left(\frac{h_{\rm max}}{h_{\rm min}}\right)+ \\ 
 \frac{\pi^2}{2}\frac{R_G^2}{h_{\rm min}^2}\left[1-\left(\frac{h_{\rm min}}{h_{\rm max}}\right)^2\right]\right\},
 \label{eq:def-F12}
\end{equation}
where $\Delta\mathcal{\tilde{F}}_{eq}$ is the \textit{ground state} equilibrium free energy difference between configurations with the center of mass either located at channel bottlenecks, $x=\pm L/2$, or at maximum channel amplitude, $x=0$. $f_{1,2}$ are the effective forces acting on the two halves of the channel. 
As pointed out in Eq.~(\ref{eq:free-energy1b}), Eq.~(\ref{eq:def-F12}) can be rewritten as
\begin{equation}
 \Delta\mathcal{\tilde{F}}_{eq}\equiv - k_BT \Delta S-\Delta\mathcal{\tilde{F}}_{eq,\infty},
 \label{eq:DeltaF-expl}
\end{equation}
where 
\begin{equation}
\Delta\mathcal{\tilde{F}}_{eq,\infty} \equiv \frac{d-1}{\beta}\frac{\pi^2}{2}\frac{R_G^2}{h_{\rm min}^2}\left[1- \left(\frac{h_{\rm min}}{h_{\rm max}}\right)^2\right]
\label{limit_N_barrier}
\end{equation}
is the asymptotic ($N\rightarrow\infty$) value of $\Delta\mathcal{\tilde{F}}_{eq}$ and it
represents the contribution of the $N-1$ monomers to the free energy difference in addition to the ''point--like`` contribution encoded in $\Delta S$. 
Substituting Eq.~(\ref{eq:approx-free-en}) into Eq.~(\ref{eq:polymer-flux-a}) it is possible to calculate, via numerical integration, the probability flux $J$ (see Appendix C) and therefore the translocation velocity $v$.

In the asymptotic regime ($N\rightarrow \infty$) we have $\Delta\mathcal{\tilde{F}}_{eq,\infty}\gg k_BT\Delta S$ and therefore we can disregard the ``point--like'' contribution $k_BT\Delta S$ to $\Delta\mathcal{\tilde F}_{eq}$. By approximating $\Delta\mathcal{\tilde F}_{eq}\simeq\Delta\mathcal{\tilde F}_{eq,\infty}$ the expression for $v$ can be simplified. In particular, we can identify two distinct scenarios. For $Nf_0L/2>\Delta\mathcal{\tilde{F}}_{eq,\infty}$, we have $f_1>f_0>f_2>0$ and the effective forces $f_{1,2}$ (see Eq.~(\ref{eq:approx-free-en})) have the same sign of the external force $f_0$ in both halves of the channel. In such a regime $v$ reads
\begin{equation}
 v\simeq \frac{D_N}{L}\left(\beta N f_0 L-2 \beta\Delta \mathcal{\tilde{F}}_{eq,\infty} \frac{\Delta \mathcal{\tilde{F}}_{eq,\infty}}{N f_0 L/2}\right).
 \label{eq:J_simpl-1}
\end{equation}
Recalling that, for a Gaussian polymer, $\Delta \mathcal{\tilde{F}}_{eq,\infty}\propto R_G^2\propto N$ and that $D_N= D_0/N$, from Eq.~(\ref{eq:J_simpl-1}) we see that, by increasing $N$, $v$ approaches to a non-vanishing constant value, consistent with our numerical results (see Fig.\ref{velocity_polymer_length_NO_SELF_f001_multipanel}). Clearly, for very large  values of the external force $N f_0 L/2 \gg \Delta \mathcal{\tilde{F}}_{eq,\infty}$ we can disregard the contribution of $\Delta \mathcal{\tilde{F}}_{eq,\infty}$ in Eq.~(\ref{eq:J_simpl-1}), leading to $v\simeq D\beta N f_0\simeq v_0$, i.e. $v$ approaches the value it attains in a constant--section channel.
On the contrary, for $\Delta\mathcal{\tilde{F}}_{eq,\infty} > N f_0 L/2$ the external work does not outnumber the free energy barrier leading to $f_1>f_0>0>f_2$ . Therefore, polymers are dragged in one of the two halves of the channel and they have to diffuse against an effective force in the other half of the channel. In this case, $v$ can be expressed as
\begin{equation}
v\simeq \frac{D_N}{L}e^{\beta f_2\frac{L}{2} }\frac{\left[\left(\beta N f_0 L\right)^2-4\left(\beta \Delta\mathcal{\tilde{F}}_{eq,\infty}\right)^2 \right]^2}{12\left(\beta \Delta\mathcal{\tilde{F}}_{eq,\infty}\right)^2+ 2\left(\beta N f_0 L\right)^2}.
\label{eq:J_simpl-2}
\end{equation}
For $N\rightarrow\infty$, we have $f_2\rightarrow-\infty$, i.e. the effective force polymers have to diffuse against diverges hence leading to vanishing small values of $v$.
According to Eqs.(\ref{eq:J_simpl-1}),(\ref{eq:J_simpl-2}) the asymptotic value of $v$ is controlled by $\frac{\Delta\mathcal{\tilde{F}}_{eq,\infty}}{Nf_{0}L/2}$: for $\frac{\Delta\mathcal{\tilde{F}}_{eq,\infty}}{Nf_{0}L/2}<1$, i.e. for an external work overcoming the free energy barrier $\Delta\mathcal{\tilde{F}}_{eq,\infty}$, the sign of the force is constant on both halves of the channel and $v$ keeps  finite, whereas for $\frac{\Delta\mathcal{\tilde{F}}_{eq,\infty}}{Nf_{0}L/2}>1$ the entropic barriers outnumbers the external force and $v$ becomes vanishing small for $N\rightarrow \infty$. 

For finite values of $N$ the scenario is more involved.  
From Eq.~(\ref{eq:approx-free-en}) we see that the sign of $f_2$ depends on the relative magnitude of $\Delta\mathcal{\tilde{F}}_{eq}$ as compared to $N f_0 L/2$. In particular, from Eqs.~(\ref{eq:approx-free-en}),(\ref{eq:def-F12}) we can identify a value of $N$ for which $f_2=0$
\begin{equation}
 N_0 \equiv \dfrac{2(d-1)\ln\left[\dfrac{h_{\rm max}}{h_{\rm min}}\right]}{\beta f_0 L -(d-1)\dfrac{\pi^2}{6}\dfrac{b^2}{h_{min}^2}\left[1-\left(\dfrac{h_{\rm min}}{h_{\rm max}}\right)^2\right]}.
 \label{eq:N_min}
\end{equation}
\begin{figure}[t]
 \includegraphics[scale=0.99]{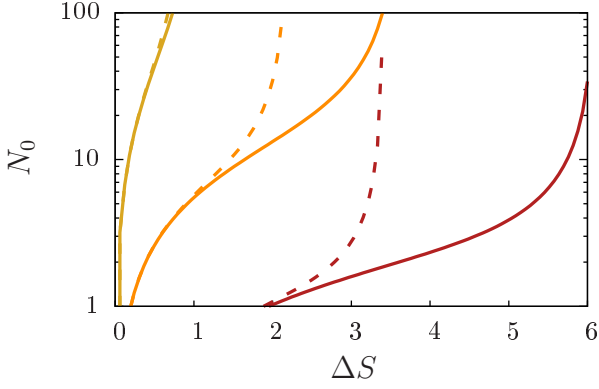}
\caption{$N_0$ (defined in Eq.~(\ref{eq:N_min})) as a function of $\Delta S$ for $\beta f_0 L=0.04,0.4,4$ for $2\mathscr{D}$ (dashed lines) and $3\mathscr{D}$ (solid lines) varying--section channels where lighter colors stand for smaller values of $\beta f_0 L$.}
\label{fig:N_min}
\end{figure}
By comparing Fig.~\ref{velocity_polymer_length_NO_SELF_f001_multipanel} with Fig.~\ref{fig:N_min} we observe that the numerical value of $N_0$, based on Eq.~(\ref{eq:N_min}) does not coincides 
with the exact value of $N$ for which the velocity attains its minimum. Nevertheless,  Eq.~(\ref{eq:N_min}) provides an insight into the non--monotonic behavior of polymer velocity. In particular, we expect a minimum in the velocity dependence on $N$ for those parameters for which $f_2(N\rightarrow \infty) > 0$ and $f_2(N=1) < 0$. In fact,  in such a regime we have $f_2<0$  for $N<N_0$ and therefore polymers have to diffuse against a force in the second half of the channel. Such a process is the rate--limiting one and since in the Rouse regime polymers diffusion coefficient decreases as $1/N$, we expect the net velocity to decrease upon increasing $N$. On the contrary, for $N>N_0$ the net force acting on polymers is positive all along the channel. Therefore, since $f_2$ grows with $N$, we expect the velocity of polymer to grow as well, until it eventually saturates for larger polymer sizes. 
Finally, Eq.~(\ref{eq:N_min}) underlines that the coupling between the external forcing and the confinement is maximized when $\frac{\Delta\mathcal{\tilde{F}}_{eq,\infty}}{Nf_{0}L/2}\simeq 1$ and vanishes for larger values of the external force.

Even capturing the qualitative behavior, the \textit{ground state} approximation does not properly predict the value $N_{min}$ at which the velocity attains its minimum. In order to have a more reliable prediction of the value $N_{min}$, at which the velocity attains its minimum, we notice that Eq.~(\ref{eq:free-energy2}) can be rewritten in dimensionless units as
\begin{multline}
\beta\mathcal{F}(x)=-\chi_1(N,f_0,L)\bar{x}-(d-1)\Biggl\{\ln\left[\frac{16\bar{h}(x)}{\pi^{2}}\right]+\\
+\ln\left[\sum_{p=1,3,..}^{\infty}\frac{1}{p^{2}}\exp\left(-\pi^{2}p^{2}\chi_2(N,h_0,\nu)^\frac{1}{\nu}\left(\frac{1}{\bar{h}(x)}\right)^{\frac{1}{\nu}}\right)\right]\Biggl\},
\label{eq:free-energy-adim}
\end{multline}
where $\bar{x}=x/L$, $\bar{h}(x)=h(x)/h_0$ and 
\begin{align}
 \chi_1(N,f_0,L)=&\beta N f_0 L \label{eq:chi1}\\
 \chi_2(N,h_0,b,\nu)=&\left(\frac{R_G(N,b,\nu)}{2h_0}\right)^\frac{1}{\nu}\label{eq:chi2}.
\end{align}
Therefore, the flux obtained from Eq.~(\ref{eq:polymer-flux-a}), as well as the probability distribution reported in  Eq.~(\ref{distr-prob-gen}) will not change if, for fixed $\Delta S$, the parameters $N$, $f_0$, $b$, $h_0$, and $L$ are varied such that $\chi_1$ and $\chi_2$ are left unchanged. Hence, for a given entropic barrier $\Delta S$ we can exploit  Eqs.~(\ref{eq:chi1}) and (\ref{eq:chi2}) to predict the value of $N_{min}$ for a new set of parameters, provided we know $\tilde{N}_{min}$  for a given set of values. For a Gaussian polymer, for which $\nu=1/2$, and $R_G=(Nb^2/6)^{1/2}$ we have
\begin{align}
 \frac{f_0 L}{\tilde{f}_0\tilde{L}}=&\frac{\tilde{h}^2_0}{h^2_0}\frac{b^2}{\tilde{b}^2}
 \label{eq:scal-2}\\
 \frac{N_{min}}{\tilde{N}_{min}}=&\frac{h^2_0}{\tilde{h}^2_0}\frac{\tilde{b}^2}{b^2}
  \label{eq:scal-1}
\end{align}
In order to check the reliability of Eqs.~(\ref{eq:scal-2}),(\ref{eq:scal-1}), we have fixed $\Delta S=2.8$ and, for constant value of $L$, we have performed numerical simulations of Gaussian polymers in $3\mathscr{D}$ varying $h_0/b$ and $f_0$. As a reference value of $N_{min}$ we have used, $\tilde{N}_{min}=10$ as obtained for $\beta \tilde{f}_0 \tilde{L}=0.4$ (see Fig.~\ref{reply1} in Appendix~\ref{app:num_est}). 
Fig.~\ref{new_fig} shows the good agreement between the prediction of $N_{min}$ provided by Eq.~(\ref{eq:scal-2}) and numerical simulations. In the appendix D, we report the numerical findings of our simulations showing the value of $N_{min}$ for different external forces.

\begin{figure}
\centering
 \includegraphics[scale=1]{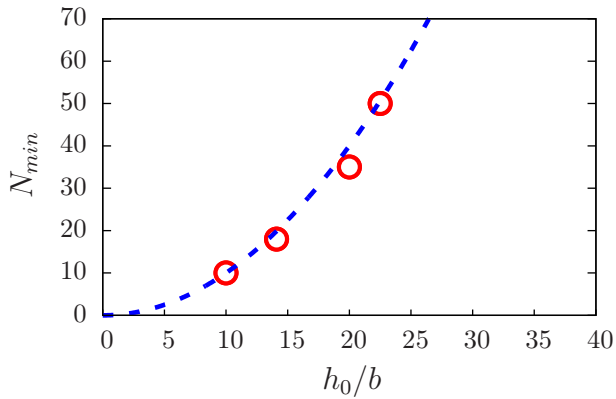}
 \caption{Polymer size $N_{min}$ at which corresponds the minimum value of the translocation velocity, normalized by $N_{min,10}=N_{min}(h_0/b=10)=9$, as a function of the channel average half-section $h_0$ and normalized by polymer monomer length $b$, as obtained from Eqs.~(\ref{eq:scal-1}) and (\ref{eq:scal-2}) (solid line) and from numerical simulations (open circles).  Data refer to the 3$\mathscr{D}$ Gaussian polymer and fixed entropic barrier $\Delta S = 2.8$.  
 Simulations have been performed with $\beta f_0 L=0.4, \text{ }0.2,\text{ }0.1, \text{ }0.08$ for $h_0=10,\text{ } 14.1,\text{ } 20,\text{ } 22.3$ respectively. Data are reported in Fig. \ref{reply} in the appendix D.}
 \label{new_fig}
\end{figure}

\section{Conclusions}
In the present work we have presented a theoretical approach aiming at describing the translocation dynamics of a polymer confined in varying--section channels.
In particular, we have shown that the translocation velocity of polymers across varying--section channels can be captured by studying the motion of a point--like particle under the action of an effective potential. Such an approach, based on the extension of the Fick--Jacobs approximation to the case of a confined polymer, is reliable when the channel section is varying on length scales bigger than the polymer linear size, i.e. its gyration radius $R_G$. We have assumed, in the spirit of the Fick--Jacobs approximation, that the effective potential experienced by the center of mass of a Gaussian polymer (Eq.~(\ref{eq:free-energy1c})) can be derived from the local equilibrium free energy of a Gaussian polymer confined between parallel plates (Eq.~(\ref{eq:doi-edw-part-funct})), separated by a distance equal to the channel section measured at the position  of the polymer center of mass.
In order to check the validity of our model we have compared our predictions with Brownian dynamics simulation for Gaussian polymers confined in a two-- ($2\mathscr{D}$) and three--dimensional ($3\mathscr{D}$) channel. 
Then, we have extended our approach to the case of  self--avoiding polymers assuming a modified expression of the free energy (Eq.~(\ref{eq:free-energy2})) that, in the limit of narrow channels, recovers the proper scaling behavior~\cite{DeGennes_book}.

We have shown that our model properly predicts the equilibrium probability distribution of the center of mass of a Gaussian as well as self--avoiding polymer (see Fig.~\ref{fig:eq-fig}). 
For both polymer models, our predictions improve for small entropic barrier $\Delta S$ (smoothly varying--section channel) and, in general, for polymers confined in three-dimensions. 
Moreover, we have shown that the predictions of our model are reliable also for non--equilibrium quantities, such as the net translocation velocity, provided that the translocation time is much larger than the slowest relaxation time of the internal modes of the polymer. Under such a condition (Eq.~(\ref{eq:def-reg-val})), our model predicts with reasonable accuracy the translocation velocity of Gaussian polymers for both $2\mathscr{D}$ and $3\mathscr{D}$ channels. In particular, for both cases we found that for smaller values of $\Delta S$ the velocity increases asymptotically upon increasing polymer size $N$,  and eventually saturates, whereas the opposite holds for larger values of $\Delta S$ for which the velocity asymptotically converges to zero. The rational of such a behavior is provided by a simplification of our model (Eq.~(\ref{eq:free-energy3})--(\ref{eq:def-F12})) 
that allows us to identify a dimensionless parameter $\frac{\Delta\mathcal{\tilde{F}}_{eq,\infty}}{Nf_{0}L/2}$, governing the translocation velocity of very long polymers ($N\rightarrow \infty$). Such a parameter represents the ratio between the asymptotic free energy barrier $\Delta\mathcal{\tilde{F}}_{eq,\infty}$, induced by the confinement, and the total work done by the external force to displace the polymer half a channel length. 
In particular, for $\frac{\Delta\mathcal{\tilde{F}}_{eq,\infty}}{Nf_{0}L/2}<1$ the external force is larger then the local entropic drift in both halves of the channel, ensuring that the effective forces $f_{1,2}$ in both haves of the channel have the same sign as the external force $f_0$.
In such a regime, for $N\rightarrow \infty$ the velocity saturates to a finite value, as shown in Eq.~(\ref{eq:J_simpl-1}). On the contrary, for $\frac{\Delta\mathcal{\tilde{F}}_{eq,\infty}}{Nf_{0}L/2}>1$ the net force on the polymer has opposite sign in the two halves of the channel. In this regime the translocation velocity is limited by the time the polymer takes to diffuse against the effective force $f_2$. Since $f_2$ increases with $N$, the translocation velocity vanishes upon increasing $N$. Interestingly, for $\frac{\Delta\mathcal{\tilde{F}}_{eq,\infty}}{Nf_{0}L/2}\lesssim 1$, we observe a non--monotonous dependence of the velocity on $N$ for both polymer models and for confinement in two and three dimensions (Fig.~\ref{velocity_polymer_length_NO_SELF_f001_multipanel}). 
The rational of such a behavior relies on the fact that the confined--induced effective force $f_2$ decreases upon increasing  $N$, as shown in Eqs.~(\ref{eq:free-energy3}),({\ref{eq:def-F12}}). Therefore we can identify a crossing polymer size $N_0$ (Eq.~(\ref{eq:N_min})), for which $f_2=0$. For $N>N_0$ we have $\Delta\mathcal{\tilde{F}}_{eq}<\beta f_0 L/2$ that leads to  $f_1>f_2>0$ and polymers experience an effective force with the same sign as the external force in both halves of the channel. On the contrary, for $\Delta\mathcal{\tilde{F}}_{eq}>\beta f_0 L/2$ we have $f_1>0>f_2$, i.e. the sign of the effective force is not constant and the net velocity of the polymer is reduced. 
Moreover, our model identifies the relevant scaling functions (Eqs.~(\ref{eq:scal-1}) and (\ref{eq:scal-2})) that control the value $N_{min}$, for which the minimum in the velocity is attained. Interestingly, our scaling functions properly predict the value of $N_{min}$ as shown in Fig.~\ref{new_fig}.

Our results can be experimentally tested, for example in micro-- and nano--fluidic devices, and can be exploited to design new devices aiming at polymer separation.

\section*{Acknowledgments}
P.M. thanks I.Pagonabarraga, T. Franosch and S. Dietrich for useful discussions. V. B. thanks L. Rovigatti and E. Locatelli for helpful discussions and suggestions. V. B. acknowledges the support from the Austrian Science Fund (FWF) project P 26253-N27.

\clearpage
\clearpage

\appendix

\onecolumngrid

\vspace{12 cm}
\section{Position dependent radius of gyration}

\begin{figure}[h]
 \includegraphics[scale=0.5]{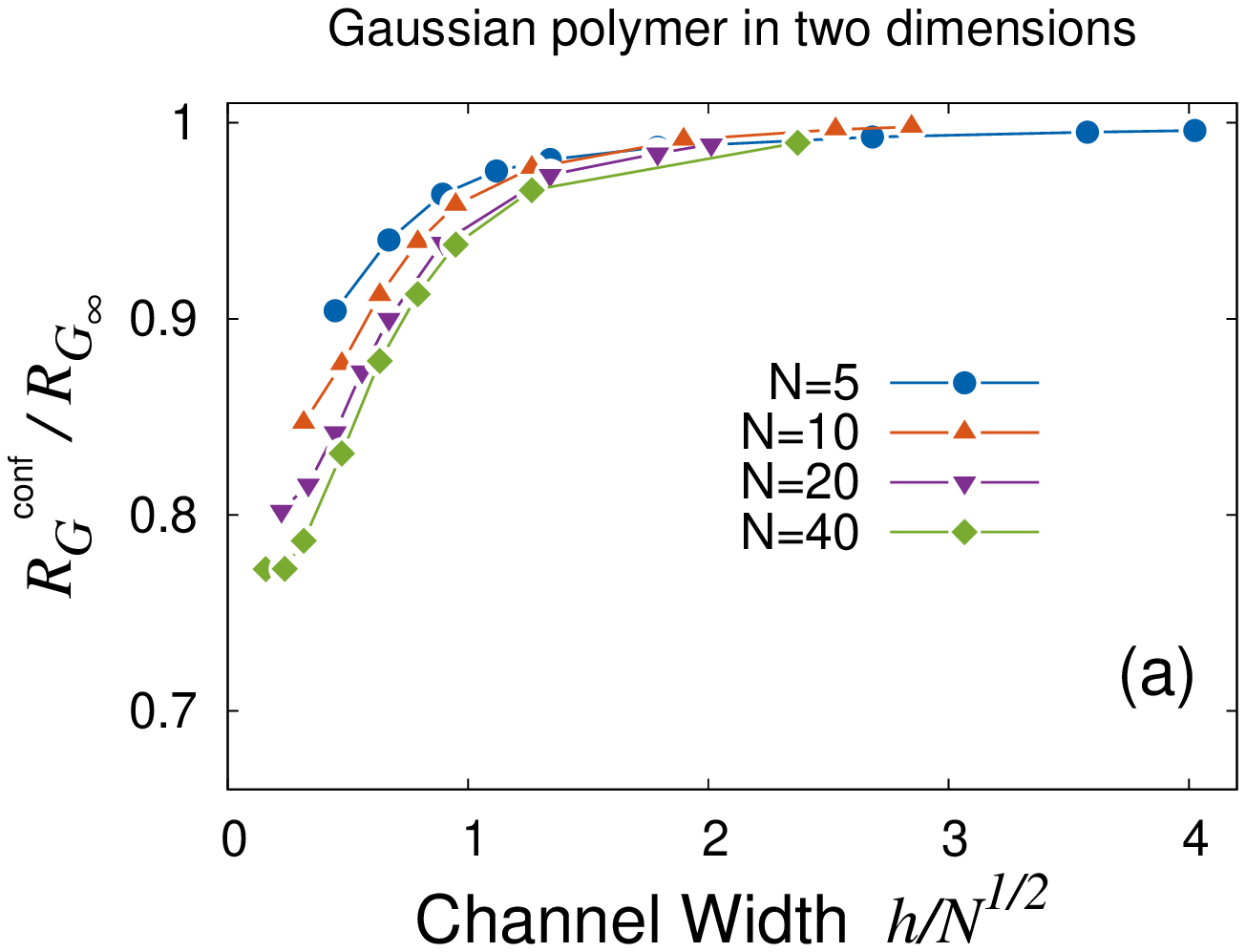}
 \includegraphics[scale=0.5]{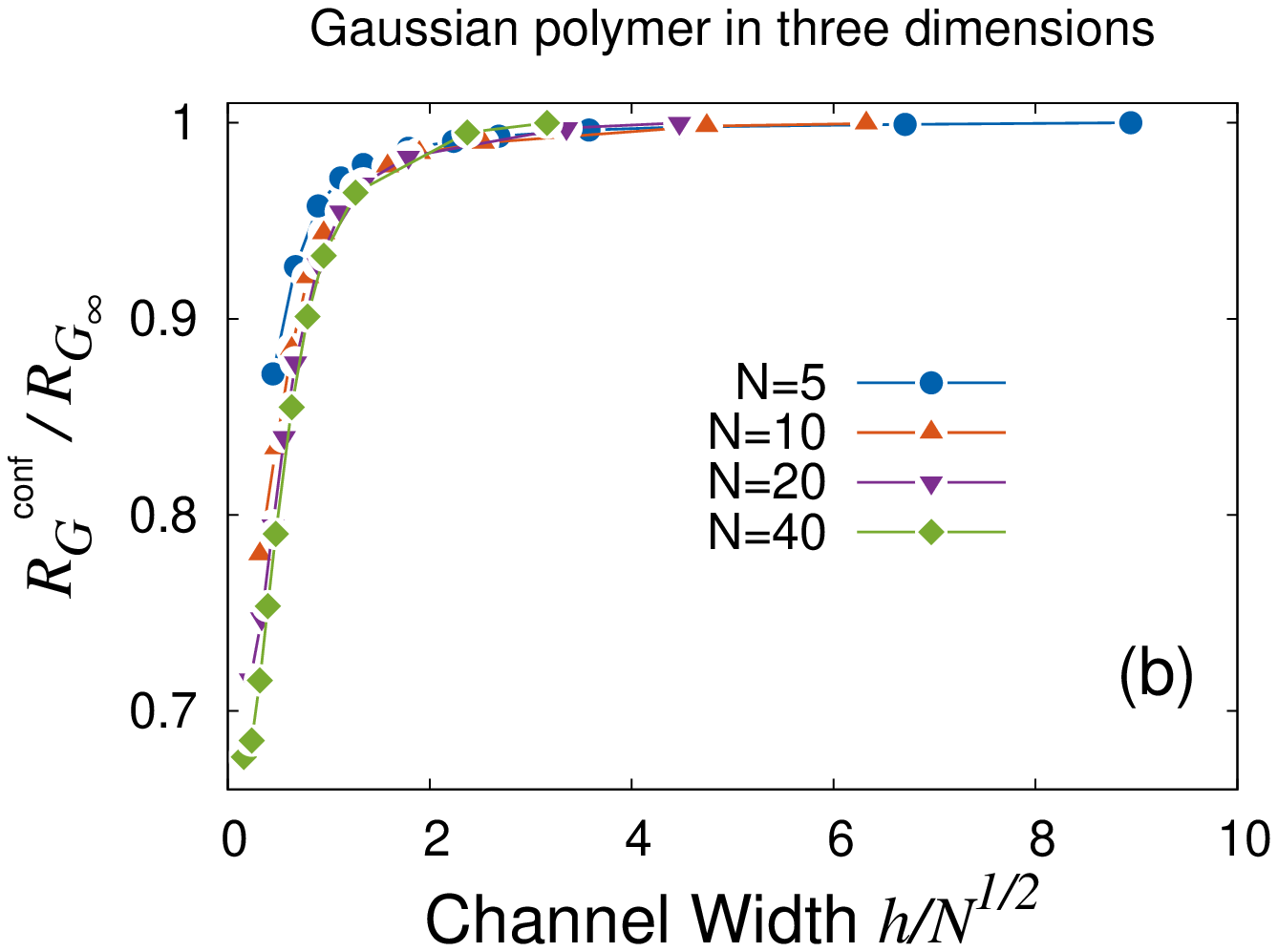}
 \vspace{-0.25 cm}
 \caption{Radius of gyration of confined Gaussian polymer $R_G^{\rm conf}$, normalized for the value assumed by $R_G$ in the bulk, $R_{G_\infty}$, as function of the polymer size normalized, for two- and three-dimensional confinement, respectively in panel (a) and (b).
  We see that $R_G^{\rm conf}$ approaches its bulk value for $h\sim N^{1/2}$, i.e. for channel width comparable with $R_{G_\infty}$. Data are calculated with Brownian dynamics simulations.}
 \label{rg_flat_channel}
\end{figure}

\begin{figure}[h]
\includegraphics[scale=0.9]{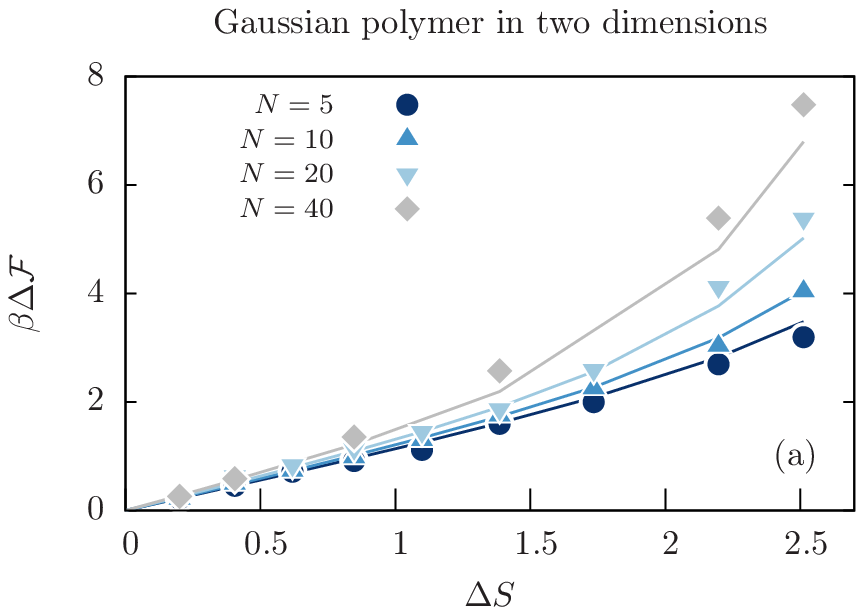}
\includegraphics[scale=0.9]{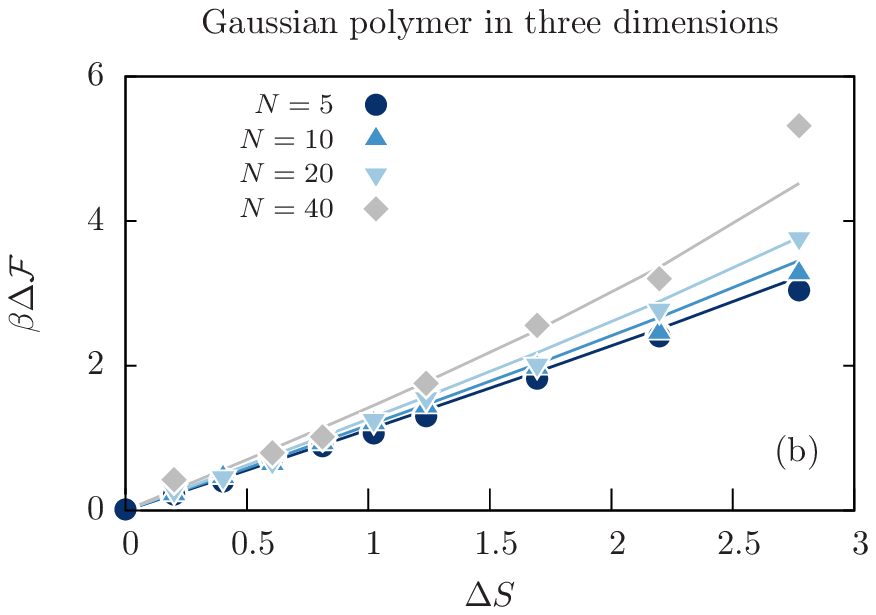}
\vspace{-0.25 cm}
\caption{Equilibrium free energy barrier between the bottleneck and the channel center for Gaussian polymers as a function of $\Delta S$ for different polymer size $N$. Data from simulations are plotted with points while lines are the theoretical predictions from Eqs.~(\ref{eq:rg_2d_flat_gauss}),(\ref{eq:free-energy1_rg})for $2\mathscr{D}$ (panel (a)) and $3\mathscr{D}$ (panel (b)) varying--section channels.}
\label{model_rg}
\end{figure}

The theoretical prediction of Eq.~(\ref{eq:free-energy1}) is based on the assumption that the gyration radius $R_G$ is constant along the channel and its value $R_{G_\infty}^2 = Nb^2/6$ corresponding the bulk value for the gyration radius of a Gaussian polymer. 
In particular, in our simulations we found that $R_G$ is not constant, rather it depends on position. 
In order to take into account such a dependence we exploit an explicit expression for the dependence of the gyration radius of a Gaussian polymer~\footnote{A different behavior is expected in the case of a self--avoiding polymer, where $R_G$, as function of channel width, $h(x)$, exhibits a non monotonic behavior that can be captured by $R_G(x)^2\sim a_2 h(x)^2 + a_3  h(x)^{-1/2}$, where $a_2$ and $a_3$ are fitting parameters~\cite{DebasishPRE}. } confined between parallel plates~\cite{DebasishPRE}
\begin{equation}
 R_G(x)^2\sim a_1 h(x)^2 + [(d-1)/d] R_{G_\infty}^2.
 \label{eq:rg_2d_flat_gauss}
\end{equation}
To find the values of the parameter $a_1$ in Eq.~(\ref{eq:rg_2d_flat_gauss}) we have fitted Eq.~(\ref{eq:rg_2d_flat_gauss}) with data from numerical simulations of Gaussian polymers confined between parallel plates (see Fig.\ref{rg_flat_channel}). 
Then have substituted Eq.~(\ref{eq:rg_2d_flat_gauss}) into Eq.~(\ref{eq:free-energy1_rg}) getting
\begin{equation}
\mathcal{F}_{eq}(x)=\frac{d-1}{\beta}\ln\Biggl[\frac{16h(x)}{h_0\pi^{2}}
\sum_{p=1,3,..}^{\infty}\frac{1}{p^{2}}\exp\left(-\frac{\pi^{2}p^{2}R_G(x)^2}{4h^{2}(x)}\right)\Biggl].
\label{eq:free-energy1_rg}
\end{equation}
Fig. \ref{model_rg} shows the enhanced agreement between the theoretical prediction and the numerical simulations once the local dependence of the gyration radius is taken into account.


\onecolumngrid

\section{Explicit approximation of the flux}

Here we report the explicit calculations leading to Eqs.~(\ref{eq:J_simpl-1}),(\ref{eq:J_simpl-2}). The starting
point is Eq.~(\ref{eq:polymer-flux-a})

\begin{align}
 J=&-D\left[\int_{-\frac{L}{2}}^{\frac{L}{2}}e^{-\beta\mathcal{F}(x)}\left(\int_{0}^{x}e^{\beta\mathcal{F}(z)}dz+\Pi\right)\right]^{-1}   \nonumber\\
 =& -D\left[\int_{-\frac{L}{2}}^{0}e^{\beta f_{1}x}\left(\int_{-\frac{L}{2}}^{x}e^{-\beta f_{1}z}dz + \Pi\right)dx +  \int_{0}^{\frac{L}{2}}e^{\beta f_{2}x}\left(\int_{-\frac{L}{2}}^{0}e^{-\beta f_{1}z}dz+\int_{0}^{x}e^{-\beta f_{2}z}dz+\Pi\right)dx\right]^{-1}  
 \\ =& -D[I_{1}+I_{2}]^{-1},  \nonumber
\end{align}
where $f_{1,2}$ are the effective forces experienced by the polymer on both the halves of the channel respectively (see Fig.~\ref{channel}.(b)). For periodic boundary conditions (see Eq.~(\ref{eq:polymer-flux-Pi})) $\Pi$ reads
\begin{equation}
\Pi=\frac{e^{\beta f_{2}\frac{L}{2}}\left(\int_{-\frac{L}{2}}^{0}e^{-\beta f_{1}x}dx+\int_{0}^{\frac{L}{2}}e^{-\beta f_{2}(x)}dx\right)}{e^{-\beta f_{1}\frac{L}{2}}-e^{\beta f_{2}\frac{L}{2}}}=\frac{\frac{1}{\beta f_{1}}\left(1-e^{-\beta f_{1}\frac{L}{2}}\right)+\frac{1}{\beta f_{2}}\left(e^{\beta f_{2}\frac{L}{2}}-1\right)}{e^{-\beta\frac{L}{2}(f_{1}+f_{2})}-1}.
\end{equation}
Concerning $I_1$ we have
\begin{equation}
I_{1}=\int_{-\frac{L}{2}}^{0}e^{-\beta f_{1}x}\left(-\frac{1}{\beta f_{1}}\left(e^{-\beta f_{1}x}-e^{\beta f_{1}\frac{L}{2}}\right)+\Pi\right)dx=-\left(\Pi+\frac{1}{\beta f_{1}}e^{\beta f_{1}\frac{L}{2}}\right)\frac{e^{\beta f_{1}\frac{L}{2}}-1}{\beta f_{1}}-\frac{1}{\beta f_{1}}\frac{L}{2}.
\end{equation}
Substituting $\Pi$ in the last expression we get
\begin{equation}
I_{1}=\frac{\frac{1}{\beta f_{1}}\left(1-e^{-\beta f_{2}\frac{L}{2}}\right)+\frac{1}{\beta f_{2}}\left(e^{-\beta f_{2}\frac{L}{2}}-1\right)}{e^{-\beta\frac{L}{2}(f_{2}+f_{1})}-1}\frac{e^{-\beta f_{1}\frac{L}{2}}-1}{\beta f_{1}}-\frac{1}{\beta f_{1}}\frac{L}{2}.
\end{equation}
The second term reads
\begin{align}
I_{2}=&\int_{0}^{\frac{L}{2}}e^{-\beta f_{2}x}\left(\frac{1}{\beta f_{1}}\left(1-e^{\beta f_{1}\frac{L}{2}}\right)+\frac{1}{\beta f_{2}}\left(e^{\beta f_{2}x}-1\right)+\Pi\right)dx \\
=& \left(\Pi+\frac{1}{\beta f_{2}}-\frac{1}{\beta f_{1}}\left(1-e^{-\beta f_{1}\frac{L}{2}}\right)\right)\frac{e^{\beta f_{2}\frac{L}{2}}-1}{\beta f_{2}}-\frac{1}{\beta f_{2}}\frac{L}{2}.
\end{align}
Substituting $\Pi$ in the last expression we get 
\begin{equation}
 I_{2}=\frac{\left[\frac{1}{\beta f_{2}}+\frac{1}{\beta f_{1}}e^{-\beta f_{1}\frac{L}{2}}\right]\left[e^{-\beta\left(f_{1}+f_{2}\right)\frac{L}{2}}-e^{-\beta f_{2}\frac{L}{2}}\right]}{e^{-\beta\left(f_{1}+f_{2}\right)\frac{L}{2}}-1}\frac{e^{\beta f_{2}\frac{L}{2}}-1}{\beta  f_{2}}-\frac{1}{\beta f_{2}}\frac{L}{2}.
\end{equation}
The value of $\Delta \mathcal{\tilde{F}}_{\infty}$ allows us to distinguish between two regimes. For $0<\Delta \mathcal{\tilde{F}}_{\infty}< Nf_0 L$ the force acting on the polymer has the same sign in both halves of the channel. Therefore, in the limit $N\rightarrow\infty$  we get
\begin{eqnarray}
 I_1&\simeq& -\frac{1}{\beta f_1}\frac{L}{2},\\
 I_2&\simeq& -\frac{1}{\beta f_2}\frac{L}{2}
\end{eqnarray}
and therefore we get 
\begin{equation}
 J\simeq \frac{D_N}{L^2}\left(\beta N f_0 L-2\beta\Delta \mathcal{\tilde{F}}_{\infty} \frac{\Delta \mathcal{\tilde{F}}_{\infty}}{N f_0 L/2}\right).
\end{equation}
In contrast, when $0<Nf_0 L<\Delta \mathcal{\tilde{F}}_{\infty}$, the total force acting on the polymer change sign at $x=0$. In this regime we have  
\begin{align}
 I_1\simeq &e^{-\beta f_{2}\frac{L}{2}}\left(\frac{\beta f_{1}-\beta f_{2}}{\beta f_{1}\beta f_{2}} \right)\frac{1}{\beta f_{2}},\\
 I_2\simeq &\left(\frac{1}{\beta f_{2}} \right)^2e^{-\beta f_{2}\frac{L}{2}},
\end{align}
that leads to 
\begin{equation}
 J\simeq \frac{D_N}{L^2}e^{\beta f_{2}\frac{L}{2}}\frac{\left[\left(\beta N f_0 L\right)^2-4\left(\beta \Delta\mathcal{\tilde{F}}_\infty\right)^2 \right]^2}{12\left(\beta \Delta\mathcal{\tilde{F}}_\infty\right)^2+ 2\left(\beta N f_0 L\right)^2},
\end{equation}
that, since $f_2<0$, in the limit $N\rightarrow \infty$ leads to $J=0$.

\section{Comparison between theoretical predicted and numerically calculated translocation velocities for larger values of the external force $f_0$}

\begin{figure}[h]
\includegraphics[scale=0.99]{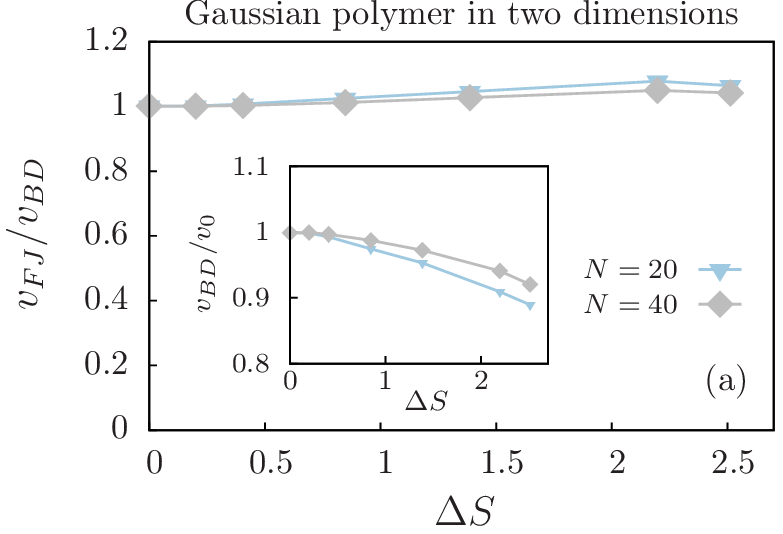}
\includegraphics[scale=0.99]{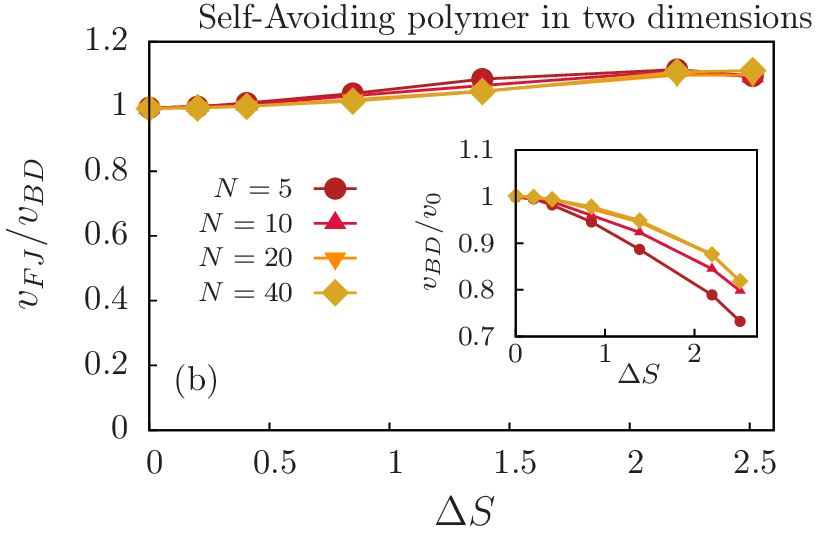}\\
\caption{Ratio between the theoretical velocity $v_{FJ}$ (calculated via Eq.~(\ref{eq:polymer-flux-a})) and the numerical velocity $v_{BD}$ (calculated via molecular dynamic simulations) as function of the entropic barrier $\Delta S$, for external force $f_0=0.1$ and in $2\mathscr{D}$ confinement.
We show the Gaussian polymer results in panel (a) and the self--avoiding polymer results in panel (b).
In the insets we show the numerical velocity, normalized for the external force, as function of $\Delta S$. }
\label{velocity_numerical_theoretical_f01}
\end{figure}

For larger values of $f_0$ the external force outnumbers the entropic contribution stemming from the confinement, hence enhancing the agreement between the theoretical prediction of the model and the translocation velocity calculated from numerical simulations (see Fig.~\ref{velocity_numerical_theoretical_f01}). 

\clearpage

\section{Numerical Estimation of $N_{min}$}\label{app:num_est}

\begin{figure}[h]
\centering
\includegraphics[scale=1]{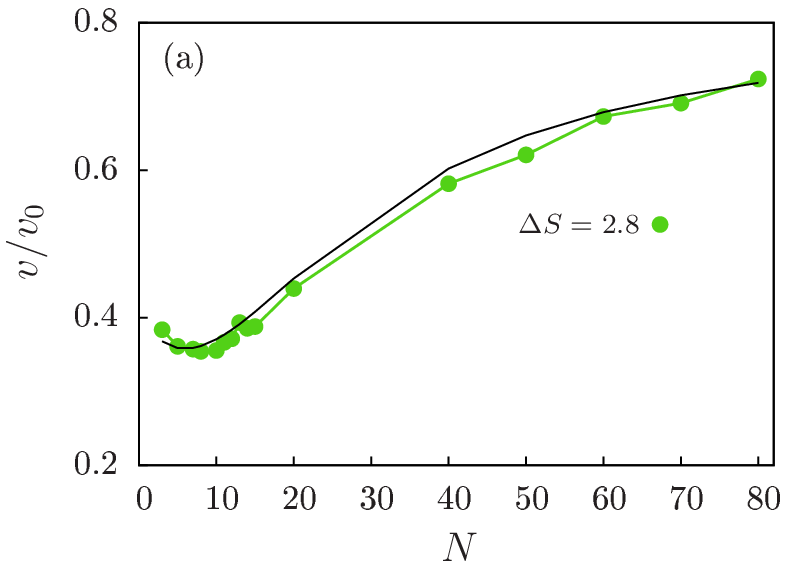}
\includegraphics[scale=1]{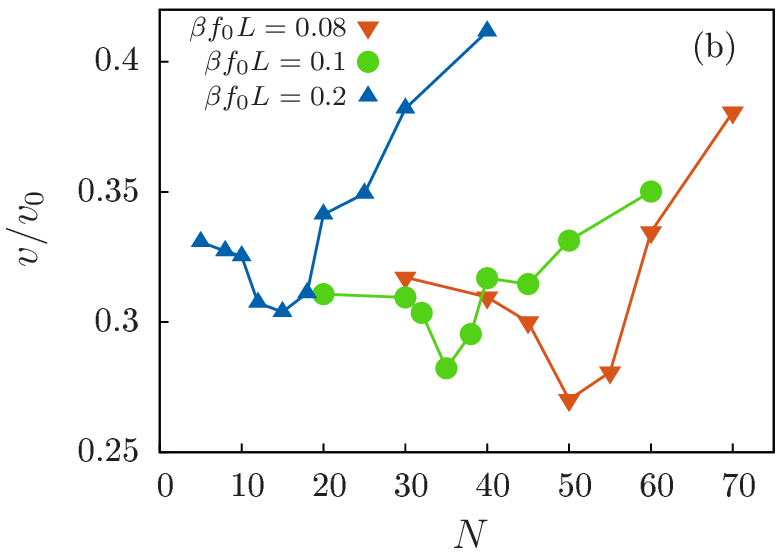}
\caption{(a) Translocation velocity for a Gaussian polymer in $3\mathscr{D}$ for a fixed entropic barrier $\Delta S =2.8$ and external forces $\beta f_0 L=0.4$. Black line represents the theoretical prediction.
(b) Translocation velocity for a Gaussian polymer in $3\mathscr{D}$ for a fixed entropic barrier $\Delta S =2.8$ and different external forces.
}
\label{reply}
\end{figure}

In Fig. \ref{reply} we report the numerical estimation of the translocation velocity for the Gaussian polymer in $3\mathscr{D}$. According to the Eq. \ref{eq:scal-1} and Eq. \ref{eq:scal-2}, decreasing $\beta f_0 L$ results in a larger value of $N_{min}$. 

\clearpage

\twocolumngrid

\end{document}